\documentclass[preprint,aps,amsmath,nofootinbib,tightenlines]{revtex4}
\usepackage{graphicx}
\usepackage{bm}
\usepackage{epsfig}
\usepackage{graphics}
\usepackage{amsfonts,amssymb}
\usepackage{color}

\def\Dsl{\hbox{/\kern-.6000em D}} 

\def\dsl{\,\raise.15ex\hbox{/}\mkern-13.5mu D}
\def\bsigma{\mbox{\boldmath $\sigma$}}

\def\psip#1{\psi_{\mathbf{#1}}}
\def\chip#1{\chi_{\mathbf{#1}}}
\def\bsigma{\mbox{\boldmath $\sigma$}}

\def\ltap{\ \raise.3ex\hbox{$<$\kern-.75em\lower1ex\hbox{$\sim$}}\ }
\def\gtap{\ \raise.3ex\hbox{$>$\kern-.75em\lower1ex\hbox{$\sim$}}\ }
\def\OMIT#1{}

\def\lsim{\mathrel{\raise.3ex\hbox{$<$\kern-.75em\lower1ex\hbox{$\sim$}}}}
\def\gsim{\mathrel{\raise.3ex\hbox{$>$\kern-.75em\lower1ex\hbox{$\sim$}}}}

\def\msb{{\overline{\rm MS}}}

\newcommand{\nn}{\nonumber}

\newcommand{\bmk}{\mathbf k}
\newcommand{\bmp}{\mathbf p}

\newcommand{\bmq}{\mathbf q}

\newcommand{\bmA}{\mathbf A}

\newcommand{\bmpp}{{\bmp^\prime}}

\newcommand{\ord}{{\cal O}}
\newcommand{\as}{\alpha_S}
\newcommand{\au}{\alpha_U}

\newcommand{\dn}{\nu \frac{d}{d \nu}}
\def\msb{{\overline{\rm MS}}}

\catcode`\@=11
\def\slash{\mathpalette\make@slash}
\def\make@slash#1#2{\setbox\z@\hbox{$#1#2$}%
  \hbox to 0pt{\hss$#1/$\hss\kern-\wd0}\box0}
\catcode`\@=12 

\newcommand{\be}{\begin{equation}}
\newcommand{\ee}{\end{equation}}
\newcommand{\bea}{\begin{eqnarray}}
\newcommand{\eea}{\end{eqnarray}}

\newcommand{\csT}{T^A \otimes {\bar T}^A}
\newcommand{\csI}{1 \otimes {\bar 1}}

\newcommand{\eps}{\epsilon}

\begin{document}


\preprint{ \vbox{ 
\hbox{UWThPh-2011-2}
\hbox{MPP-2011-8cd}  
}}

\title{\phantom{x}\vspace{0.5cm} 
Ultrasoft NLL Running of the Nonrelativistic  \\
 $O(v)$ QCD Quark Potential
\vspace{1.0cm} }

\author{Andr\'e~H.~Hoang\footnote{Electronic address: Andre.Hoang@univie.ac.at}$^{1,2}$ and Maximilian~Stahlhofen\footnote{Electronic address: Maximilian.Stahlhofen@univie.ac.at}$^{1,3}$}
\affiliation{\vspace{1 cm}
$^1$University of Vienna, Faculty of Physics, Boltzmanngasse 5, A-1090 Wien, Austria \vspace{0.5
  cm}\\
$^2$Max-Planck-Institut f\"ur Physik, F\"ohringer Ring 6, 80805 M\"unchen, Germany\vspace{0.5 cm}\\
$^3$Grup de F\'\i sica Te\`orica and IFAE, Universitat Aut\`onoma de Barcelona, E-08193 Bellaterra, Barcelona, Spain\vspace{1cm}\\
}


\begin{abstract}
\vspace{0.5cm}
\setlength\baselineskip{18pt}
Using the nonrelativistic effective field theory vNRQCD, we determine the
contribution to the next-to-leading logarithmic (NLL) 
running of the effective quark-antiquark potential at order $v$ ($1/m|{\bf k}|$)
from diagrams with one potential and two ultrasoft loops, $v$ being the velocity
of the quarks in the c.m. frame. The results are numerically important and complete
the description of ultrasoft next-to-next-to-leading logarithmic (NNLL) order
effects in heavy quark pair production and annihilation close to threshold.  
\end{abstract}
\maketitle


\newpage

\section{Introduction}
\label{sectionintroduction}

The measurement of the line-shape of the total top-antitop quark
production cross section in the threshold  
region $\sqrt{s}\approx 2m_t$ is one of the major tasks of the top quark
physics program at a future linear collider. The most prominent quantity to be
measured is the top quark mass, and one can expect an improvement in precision of $m_t$ by 
about an order of magnitude to mass measurements based on reconstruction
obtained at the Tevatron and the LHC~\cite{thresholdscan}. In addition, fitting
the lineshape measurements to theoretical higher order predictions allows to
control precisely the top mass scheme, which is not the case for mass
reconstruction at hadron colliders based on Monte-Carlo generators.

To obtain a meaningful theoretical description of the nonrelativistic
threshold dynamics it is required to systematically sum the so-called Coulomb
singular terms $\propto (\alpha_s/v)^n$ in a systematic
nonrelativistic expansion, $v$ being the relative velocity of the top quarks. This 
task is achieved by means of effective theories based on nonrelativistic QCD
(NRQCD)~\cite{BBL}. In this approach, however, sizeable logarithmic terms
$\propto(\alpha_s\ln v)^n$ are not systematically accounted for, which leads
to rather large normalization uncertainties of the cross section
line-shape. The next-to-next-to-leading order (NNLO) predictions in this
``fixed-order'' approach were estimated to have a normalization uncertainty of
order $20\%$~\cite{synopsis}. Presently known NNNLO fixed-order corrections, e.g.~\cite{Kniehl:2002yv,Beneke:2008cr,Anzai:2009tm,Smirnov:2009fh}, seem to reduce this uncertainty to about 10\%~\cite{Beneke:2008ec}.
This normalization uncertainty might not affect the top mass measurement, which primarily depends on the c.m. energy where the cross section rises, but it still renders precision measurements of other quantities such as the top total width or the top quark couplings impossible. On the other hand, such large normalization uncertainties might cast general doubts on the reliability of the theoretical method itself.
To match the statistical uncertainties, that are expected for these quantities at an
International Linear Collider (ILC), a theoretical precision of the cross
section normalization of at least $3\%$ would be required.

In Refs.~\cite{HMST} (see also Ref.~\cite{PinedaSigner}) it was
demonstrated that the 
summation of logarithmic $(\alpha_s\ln v)^n$ terms, using
renormalization-group-improved (RGI) perturbation theory, can significantly reduce the
normalization uncertainties of the threshold cross section. 
Concerning QCD effects the RGI leading-logarithmic (LL) and NLL order predictions of the threshold cross section
are completely known, but no full NNLL order prediction exists at present. 
The full NNLL order prediction is, however, required to obtain a
reliable estimate of the remaining theoretical normalization uncertainties. 

The missing ingredient is the NNLL running of the Wilson coefficient of the
leading order effective current that describes production and annihilation of
a nonrelativistic $t\bar t$ pair in a S-wave spin-triplet state
(${}^3S_1$). Adopting the label notation from
vNRQCD~\cite{LMR,HoangStewartultra} the current has the form
\begin{align} 
  {\bf J}_{1,\bf p} = 
    \psi_{\bmp}^\dagger\, \bsigma (i\sigma_2) \chi_{-\bmp}^*
\,,
\label{J1J0}
\end{align}
where  $\psi_{\bmp}$ and $\chi_{\bmp}$ annihilate top and antitop quarks with
three-momentum $\bmp$, respectively, and where color indices have been
suppressed. The current does not have a LL anomalous dimension because there
is no one-loop vertex diagram in the effective theory that contains UV
divergences associated with the current ${\bf J}_{1,\bf p}$. Such UV
divergences arise at NLL order from insertions of the next-to-next-to-leading kinetic energy
operators and from insertions of the next-to-next-to-leading order potentials~\cite{LMR}. The
corresponding computations were carried out in
Refs.~\cite{amis,Pineda:2001et,HoangStewartultra} and are 
completed~\cite{Pineda:2001et,HoangStewartultra}. Using the conventions
from~\cite{HoangStewartultra} the 
resulting NLL order renormalization group equation (RGE) for the Wilson
coefficient $c_1$ of the current ${\bf J}_{1,\bf p}$ has the form (${\bf S}^2=2$)
\begin{align}
 \left(\nu \frac{\partial}{\partial\nu} \ln[c_1(\nu)] \right)^{\rm NLL} 
\!\!\!\! =  
 -\:\frac{{\cal V}_c^{(s)}(\nu)
  }{ 16\pi^2} \bigg[ \frac{ {\cal V}_c^{(s)}(\nu) }{4 }
  +{\cal V}_2^{(s)}(\nu)+{\cal V}_r^{(s)}(\nu)
   + {\bf S}^2\: {\cal V}_s^{(s)}(\nu)  \bigg] 
 +\frac12 {\cal V}_{k,\rm eff}^{(s)}(\nu)\,,
\label{c1anomdim}
\end{align}
where $\nu$ is the vNRQCD velocity renormalization parameter that is
conveniently used to parametrize the correlation between soft and ultrasoft
dynamical scales within the renormalized effective
theory~\cite{LMR}. The generalization of Eq.~\eqref{c1anomdim} for currents
describing pairs of quarks and colored scalars in any angular momentum and spin
state (${}^{2s+1}L_J$) were derived in Ref.~\cite{HoangRuiz2}.

In Eq.~(\ref{c1anomdim}) the term ${\cal V}_c^{(s)}$ is the Wilson
coefficient of the 
Coulomb potential $\propto 1/\bmk^2$, and ${\cal V}_2^{(s)}$ and ${\cal
  V}_r^{(s)}$ are the coefficients of the ${\cal O}(v^2)$ potentials with the
momentum structure $1/m^2$ and $(\bmp^2+{\bmp^{\prime}}^2)/(2m^2\bmk^2)$,
respectively, $m$ being the heavy quark mass. The term ${\cal V}_s^{(s)}$
is the coefficient of the spin-dependent potential that can contribute for
spin triplet S-wave states. The coefficient ${\cal V}_{k,\rm eff}^{(s)}$
combines Wilson coefficients from several non-Abelian  $1/m|\bmk|$
potentials at ${\cal O}(v)$. At LL order it has the form 
\begin{eqnarray}
 {\cal V}_{k, \rm eff}^{(s)}(\nu) &=& 
\alpha_s(m \nu)^2 \, \frac{C_F}{2}(C_F-2 C_A) 
\,+\, \big({\cal V}_{k, \rm eff}^{(s)}(\nu)\big)_{us}\;.
�
\label{Vkeffinc1}
\end{eqnarray}
The first term arises from soft renormalization and the second 
represents the evolution from ultrasoft UV divergences.\footnote{
There is a misprint in Eq.(4) of Ref.~\cite{3loop} where a factor $1/2$ is
missing for the soft contributions. All numerical computations in Ref.~\cite{3loop} were carried out
with the correct result.} 
The terms in Eq.~(\ref{Vkeffinc1}) are defined such that the ultrasoft terms
have zero matching condition at $\nu=1$, $({\cal V}_{k, \rm
  eff}^{(s)}(1))_{us}=0$. The superscripts ${(s)}$ refer to the color singlet 
state of the quark pair. 

At NNLL order there are two types of contributions that have to be considered
for the evolution of $c_1$. The first arises from three-loop vertex diagrams
that come from insertions of subleading soft matrix element corrections to the
potentials and from insertions of potentials with additional exchange of
ultrasoft gluons. The corresponding results were determined in
Ref.~\cite{3loop} and are referred to as the {\it non-mixing} contributions
as they affect the evolution of $c_1$ directly through UV-divergences. The
second type of contributions arises from the subleading evolution of the
potential Wilson coefficients that appear in the NLL order RGE shown in Eq.~(\ref{c1anomdim}). They are
referred to as the {\it mixing} contributions as they affect the evolution of $c_1$
indirectly. Except for the coefficient of the Coulomb
potential ${\cal V}_c^{(s)}$~\cite{Pineda:2001ra,HoangStewartultra} and for
the spin-dependent potential ${\cal V}_s^{(s)}$~\cite{Penin:2004xi} no
complete determination for the subleading evolution 
exists at present. 

The analysis of the three-loop (non-mixing) terms in Ref.~\cite{3loop} showed
that the contributions 
involving the exchange of ultrasoft gluons are more than an order of magnitude
larger than those arising from soft matrix element insertions and in fact
similar in size to the previously known NLL contributions. The reason is
related to the larger size of the ultrasoft coupling $\alpha_s(m \nu^2)$ and to
a rather large coefficient multiplying the ultrasoft contributions. These large
ultrasoft contributions are responsible for an uncertainty in the normalization
of the most up-to-date top pair threshold cross section prediction of at best 
$6\%$~\cite{HoangEpiphany}, which is quite far from the required precision
(see also Ref.~\cite{PinedaSigner}). From this analysis it is reasonable to
assume that the ultrasoft effects, which form a gauge-invariant subset, also
dominate the mixing contributions. This is also consistent with the small
numerical effects~\cite{Penin:2004ay} of the NLL evolution of the spin-dependent
coefficient ${\cal V}_s^{(s)}$~\cite{Penin:2004xi}, which is dominated by soft
effects and receives ultrasoft contributions only indirectly through
mixing.

In Ref.~\cite{V2Vr} we have computed the ultrasoft contribution to the
NLL anomalous dimensions of the spin-independent $1/m^2$ potential coefficients 
${\cal V}_2^{(s)}$ and ${\cal V}_r^{(s)}$. Here UV-divergent diagrams with two
ultrasoft loops contribute directly to their anomalous dimensions. These NLL
ultrasoft corrections were indeed large and found to be similar in size to the
previously known LL order evolution. Their overall contribution in the NNLL
mixing corrections to the NLL RGE in
Eq.~(\ref{c1anomdim}), however, turned out to be quite small due to the
$1/16\pi^2$ suppression factor. 

In this work we complete the determination of the NNLL ultrasoft corrections to
the anomalous dimension of $c_1$ by the computation of the NLL ultrasoft
renormalization group (RG) evolution of the coefficients 
$({\cal V}_{k,\rm eff}^{(s)})_{us}$. Since its contribution in
Eq.~(\ref{c1anomdim}) is substantially larger than those of the 
${\cal V}_{2,r}^{(s)}$, the impact of the NLL ultrasoft corrections can be
expected to be sizeable and might compensate the large ultrasoft NNLL non-mixing
corrections to $c_1$ determined in Ref.~\cite{3loop}. As we show in this work,
this is indeed the case as the results of this work substantially stabilize the
RG evolution of $c_1$. 
Our result will contribute to a reduction of the theoretical uncertainty of
current RGI predictions for the heavy quark pair
threshold production rate and in particular for top quark pair production at a
future linear collider.

This work is organized as follows:
In Sec.~\ref{sectiontheory} we briefly review the effective theory setup used for our work and define the relevant operators in the vNRQCD Lagragian. The ultrasoft renormalization procedure with regard to the $1/m|\bmk|$ potentials is explained in Sec.~\ref{sectionrenproc}. Section~\ref{sectionLLresults} gives a short overview of the respective calculation at the LL level in Feynman gauge, before we present the new calculation at NLL order in Sec.~\ref{sectionCalc}. In Sec.~\ref{secresults} we summarize the results for the $1/m|\bmk|$ and $1/m^2$ potentials and present the complete ultrasoft NNLL mixing contribution to the running of the current coefficient $c_1$. Sec.~\ref{sectiondiscussion} contains a numerical analysis of our results and Sec.~\ref{sectiondiscussion} our conclusion.

\section{Theoretical Setup}
\label{sectiontheory}

The vNRQCD Lagrangian is organized as an expansion in the heavy quark velocity
$v$. It consists of three mayor parts
\cite{LMR,HoangStewartultra,ManoharVk}, 
\begin{equation}
 {\cal L}_{\rm vNRQCD}\:=\:  {\cal L}_{u} \:+\:  {\cal L}_{p} \:+\:  {\cal L}_{s}
\,,
\end{equation}
containing kinetic terms and ultrasoft interactions (${\cal L}_{u}$), potential
interactions (${\cal L}_{p}$) 
and interactions involving soft degrees of freedom (${\cal L}_{s}$), respectively. The
ultrasoft term ${\cal L}_u$  has the form
\begin{equation}
{\cal L}_u  = 
\sum_{\mathbf p} \bigg\{
   \psi_{\bmp}^\dagger   \!\bigg[ i D^0 - \frac {\left({\bf p}\!-\!i{\bf D}\right)^2}
   {2 m}
 + \frac{\bmp^4}{8 m^3}
 + \ldots \!\bigg]\! \psi_{\bmp}
 + (\psi \!\to\! \chi,\, T \!\to\! \bar T) \bigg\}\!
 -\frac{1}{4}G^{\mu\nu}G_{\mu \nu} 
+\ldots ,\!
\label{Lus}
\end{equation}
where $D^\mu = \partial^\mu + i g_U A^\mu(x)$ is the ultrasoft gauge-covariant
derivative, $g_U$ is the ultrasoft coupling constant, and $G^{\mu\nu}$ is the
ultrasoft field strength tensor. Besides the propagation of  
the heavy quarks ${\cal L}_u$ describes their interaction with ultrasoft gluons.

The potential term ${\cal L}_p$ contains the potential interactions
between the quark and the antiquark and can be written as 
\be
{\cal L}_p = {\cal L}_{pV} +  {\cal L}_{pu} + {\cal L}_{pk}
\,,
\ee
where   
\begin{eqnarray}
{\cal L}_{pV}&=& - \sum_{\bmp, \bmpp}  
V_{\alpha\beta\lambda\tau} \left({\bmp},{\bmpp}\right)\ 
\psi_{\bmpp \alpha}^\dagger\: \psi_{\bmp\, \beta}\:
  \chi_{-\bmpp \lambda}^\dagger\:  \chi_{-\bmp\, \tau} + \ldots  
\label{Lpot}
\end{eqnarray}
consists of potential four-quark operators. They depend on the soft three-momentum
labels $\bmp$ and $\bmp^\prime$ of the heavy quark fields, and the 
coefficients have the form
\begin{eqnarray}
V_{\alpha\beta\lambda\tau}({\bmp},{\bmp^\prime}) &=& 
(T^A_{\alpha\beta} \otimes \bar T^A_{\lambda\tau})\, \bigg[
 \frac{{\cal V}_c^{(T)}}{\bmk^2}
 + \frac{{\cal V}_r^{(T)}({\bmp^2 + \bmp^{\prime 2}})}{2 m^2 \bmk^2}
 + \frac{{\cal V}_2^{(T)}}{m^2}
 + \ldots \bigg] \nn \\
&& + \,(1_{\alpha\beta}\otimes \bar 1_{\lambda\tau})\, \bigg[
 \frac{{\cal V}_c^{(1)}}{\bmk^2}
 + \frac{{\cal V}_r^{(1)}({\bmp^2 + \bmp^{\prime 2}})}{2 m^2 \bmk^2}
 + \frac{{\cal V}_2^{(1)}}{m^2}
 +\ldots \bigg]
\,,
\label{pots}
\end{eqnarray}
the terms
$\alpha,\beta,\lambda,\tau$ being color indices and $\bmk \equiv \bmpp-\bmp$. 
The ellipses in Eq.~\eqref{pots} refer to spin-dependent ${\cal O}(v^2)$
potentials, that are not relevant in this work, as well as to higher orders in
the $v$ expansion. 
The leading order terms in Eq.~\eqref{pots} in the velocity power counting are the
Coulomb potential operators. 
Their LL coefficients are ${\cal V}_c^{(T)}(\nu)=4 \pi \alpha_s(m
\nu)$ and ${\cal V}_c^{(1)}(\nu)=0$, where $\nu$ is the vNRQCD velocity
renormalization scaling parameter~\cite{LMR}.

The term ${\cal L}_{pu}$ includes the higher order terms in the multipole
expansion related to the potentials in Eq.~\eqref{pots}. It is fixed by
reparametrization and (ultrasoft) gauge invariance of ${\cal
  L}_p$~\cite{ManoharVk}: 
\begin{eqnarray}
  {\cal L}_{pu} &=& \frac{ 2 i\: {\cal V}_c^{(T)}\, f^{ABC} }{ {\bmk}^4}\, 
    {\bmk}\cdot (g {\bmA}^C) \: \psip{p^\prime}^\dagger\:
  T^A {\psip p}\: \chip{-p^\prime}^\dagger\: \bar T^B {\chip {-p}}{}+  \label{Lpu} \\[1.5 ex]
  && +\: {\cal V}_c^{(T)} 
 \psip{p^\prime}^\dagger\:
    \bigg[ \frac{i{\bmk}\cdot
    \overleftrightarrow{\nabla}}{{\bmk}^4} -\frac{\overleftrightarrow{\nabla}^2}{2{\bmk}^4}
    +2 \frac{({\bmk}\cdot \overleftrightarrow{\nabla})^2}{{\bmk}^6} \bigg]
  T^A {\psip p}\:\chip{-p^\prime}^\dagger\: \bar T^A {\chip {-p}}{}+ \nn\\[1.5 ex]
  && +\: {\cal V}_c^{(T)} 
 \psip{p^\prime}^\dagger\:
  T^A {\psip p}\:
  \chip{-p^\prime}^\dagger\: \bigg[ \frac{-i{\bmk}\cdot
    \overleftrightarrow{\nabla}}{{\bmk}^4} -\frac{\overleftrightarrow{\nabla}^2}{2{\bmk}^4}
    +2 \frac{({\bmk}\cdot \overleftrightarrow{\nabla})^2}{{\bmk}^6} \bigg]
    \bar T^A {\chip {-p}}{} + \;\ldots\;. \qquad \nn
\end{eqnarray}
The first term in Eq.~\eqref{Lpu} describes the coupling of an ultrasoft gluon
to the Coulomb potential,  
the other terms are four-quark operators with ultrasoft derivatives ${\bf
  \overleftrightarrow{\nabla} = \overrightarrow{\nabla} +
  \overleftarrow{\nabla}}$ acting on the fermion fields to the left and to the
right. 
The associated Feynman rules can be found in App.~\ref{feynmanrules}.

In contrast to early works~\cite{LMR,amis,ManoharVk}, we omit potential terms of
the form 
\be
(T^A_{\alpha\beta} \otimes \bar T^A_{\lambda\tau})\, 
 \frac{{\cal V}_k^{(T)}\pi^2}{m|{\bmk}|}
 + \,(1_{\alpha\beta}\otimes \bar 1_{\lambda\tau})\,
 \frac{{\cal V}_k^{(1)}\pi^2}{m|{\bmk}|}
\label{4qVk}
\ee
in Eq.~\eqref{pots}. 
In the term 
\be
{\cal L}_{pk} = \sum_{i, X} {\cal V}_{ki}^{(X)}\,{\cal O}_{ki}^{(X)}
\,+\,\ldots
\,,
\label{L6Q}
\ee
we instead introduce the sum operators ${\cal O}_{ki}$ as they are suitable to carry out the
ultrasoft renormalization 
procedure~\cite{HoangStewartultra,3loop}. These sum operators generate a 
potential $V_{k} \propto 1/m ({\bmk}^2)^{\frac{5-d}{2}}$ upon summation 
of the intermediate soft label momentum $\bmq$, when inserted into vNRQCD matrix
elements.
The ellipses in Eq.~\eqref{L6Q} indicate other sum
operators~\cite{HoangStewartultra,3loop} 
which we, however, do not need to consider in this work.
For the sum operators needed in our work we use the following operator basis,
$\mu_S=m\nu$ being the soft renormalization scale:
\begin{align}
{\cal O}_{k1}^{(1)} =& -\frac{
[{\cal V}_c^{(T)}(\nu)]^2\mu_S^{4\eps}}{m}\,\csI
 \sum_{\bmp,\bmpp,\bmq} ( f_{0} + f_{1} + 2 f_{2} )\
 \big[ \psi_{\bmpp}^\dagger \,\psi_{\bmp}\,
 \chi_{-\bmpp}^\dagger \, \chi_{-\bmp} \big] \,, \nn\\
{\cal O}_{k2}^{(T)} =& -\frac{
[{\cal V}_c^{(T)}(\nu)]^2\mu_S^{4\eps}}{m}\,\csT
 \sum_{\bmp,\bmpp,\bmq} ( f_{1} + f_{2} ) \ 
 \big[ \psi_{\bmpp}^\dagger \, \psi_{\bmp}\,
 \chi_{-\bmpp}^\dagger \, \chi_{-\bmp} \big] \,,  \nn\\
{\cal O}_{k3}^{(1)} =& -\frac{
[{\cal V}_c^{(T)}(\nu)]^2\mu_S^{4\eps}}{m}\,\csI
 \sum_{\bmp,\bmpp,\bmq} (f_0 + f_1) \ 
 \big[ \psi_{\bmpp}^\dagger \, \psi_{\bmp}\,
 \chi_{-\bmpp}^\dagger \, \chi_{-\bmp} \big] \,.  \nn\\
{\cal O}_{k3}^{(T)} =& -\frac{
[{\cal V}_c^{(T)}(\nu)]^2\mu_S^{4\eps}}{m}\,\csT
 \sum_{\bmp,\bmpp,\bmq} (f_0 + f_1) \ 
 \big[ \psi_{\bmpp}^\dagger \, \psi_{\bmp}\,
 \chi_{-\bmpp}^\dagger \, \chi_{-\bmp} \big] \,. \label{Oki}
\end{align}
The functions $f_i$ depend on the external soft three-momentum labels $\bmp$, $\bmpp$
and the intermediate soft three-momentum label $\bmq$. 
They are defined as~\cite{HoangStewartultra} 
\begin{eqnarray}
 f_0 &=& \frac{\bmp^\prime\cdot (\bmq-\bmp)}{(\bmq-\bmp)^4\,
  (\bmq-\bmp^\prime)^2} 
  + (\bmp \leftrightarrow \bmp^\prime)\,, \nn\\[2 mm]
 f_1 &=& \frac{\bmq \cdot (\bmq-\bmp)}{(\bmq-\bmp)^4\,(\bmq-\bmp^\prime)^2} 
  + (\bmp \leftrightarrow \bmp^\prime)\,, \label{ffunctions} \\[2 mm]
 f_2 &=& \frac{(\bmq-\bmp^\prime)\cdot (\bmq-\bmp)}
  {(\bmq-\bmp)^4\,(\bmq-\bmp^\prime)^4}\: (\bmq^2-\frac{\bmp^{\prime\,2}}{2}-\frac{\bmp^2}{2})\;.\nn
\end{eqnarray}
In four-quark matrix elements 
the sum over $\bmq$ is understood to be replaced by a soft loop integral~\cite{LMR}, i.e.
\begin{eqnarray}
 \sum_{\bmq} \to \int\!\! \frac{d^{d-1}q}{(2\pi)^{d-1}}  
\,.
\label{sum}
\end{eqnarray}
The $\bmq$ integrals over the $f_i$'s in $d-1$ spatial dimensions yield
functions that only depend on $\bmk = \bmpp-\bmp$,
\begin{align}
\int\!\!\! \frac{d^{d-1}q}{(2 \pi)^{d-1}}\,f_0 &= \frac12 |\bmk|^{d-5} \Big[ f(1,1)+f(1,2) \Big]\nn
 \;\xrightarrow{d\to4}\; \frac1{16 |\bmk|}\,,\\
\int\!\!\! \frac{d^{d-1}q}{(2 \pi)^{d-1}}\,f_1 &= \frac12 |\bmk|^{d-5} \Big[ 3 f(1,1) - f(1,2) \Big]\nn
 \;\xrightarrow{d\to4}\; \frac3{16 |\bmk|}\,,\\
\int\!\!\! \frac{d^{d-1}q}{(2 \pi)^{d-1}}\,f_2 &= \frac14 |\bmk|^{d-5} \Big[ 2 f(1,1) - 4 f(1,2) + f(2,2) \Big]
 \;\xrightarrow{d\to4}\; \frac1{16 |\bmk|}\,,
\label{intfs}
\end{align}
where
\begin{align}
f(a,b)=\frac{\Gamma \left(a+b-\frac{d-1}{2}\right) \Gamma \left(\frac{d-1}{2}-a\right) \Gamma \left(\frac{d-1}{2}-b\right) }{\Gamma (a) \Gamma (b) \Gamma (d-1-a-b) (4 \pi)^{\frac{d-1}{2}}}\,.
\end{align}
They reproduce for $d\to 4$ the $1/|\bmk|$ potentials mentioned in
Eq.~(\ref{4qVk}). For simplicity we call the sum operators ${\cal
  O}_{k,i}^{(X)}$ also $V_k$ potentials in the rest of this work. 
The operators ${\cal O}_{k1}^{(1)}$ and ${\cal O}_{k2}^{(T)}$ have been
introduced before in Ref.~\cite{HoangStewartultra} for the LL ultrasoft
renormalization. The operators  ${\cal O}_{k3}^{(1,T)}$ are new and required for
the presentation of our computations at NLL order. 

Unlike the corresponding operators in Eq.~\eqref{4qVk} the sum operators can be
renormalized consistently beyond one-loop level~\cite{HoangStewartultra,PhD}. By
including the renormalized factor $[{\cal V}_c^{(T)}(\nu)]^2$ in the definition
of the ${\cal O}_{ki}$ we anticipate a factorization of the soft (LL)
contributions (encoded 
in ${\cal V}_c^{(T)}(\nu)$) and the ultrasoft contributions (encoded in ${\cal
  V}_{ki}^{(X)}(\nu)$) to their RG running. An analogous factorization occurs
for the $1/m^2$ potentials ${\cal V}_r$ and ${\cal V}_2$
~\cite{HoangStewartultra,V2Vr}.\footnote{
A complete account on how this factorization arises requires the introduction of additional sum operators with four heavy quark and two soft fields, which, analogous to the ${\cal O}_{ki}$, absorb the UV divergences of vNRQCD diagrams with external soft fields and ultrasoft and potential loops.
By forming a soft tadpole these operators contribute soft mixing
terms to the anomalous dimension of the ${\cal O}_{ki}$. Together with the
contributions in this work they produce the factorized running of the
coefficient $[{\cal V}_c^{(T)}(\nu)]^2\, {\cal V}_{ki}^{(X)}(\nu)$. This
approach has been adopted in Ref.~\cite{V2Vr}.
We do not follow this approach here and refer to Ref.~\cite{PhD} for a detailed 
discussion.  
In the two-stage matching approach adopted in ``potential'' NRQCD
(pNRQCD)~\cite{Pineda:1997bj}, this kind of factorization in the potential
coefficients is implemented by construction.}
In this paper we compute the (ultrasoft) NLL running of the 
${\cal V}_{ki}(\nu)$. The LL running was determined in
Ref.~\cite{HoangStewartultra} in Coulomb gauge.  

Using the equation
\begin{eqnarray}
 \left[\begin{array}{c} V_{\rm singlet} \cr V_{\rm octet} \end{array}\right]
 =\left[\begin{array}{ccc} 1 &  & -C_F \cr
    1 &  & \frac{1}{2} C_A - C_F \cr
 \end{array}\right]
 \left[\begin{array}{c} V_{1\otimes 1} \cr V_{T\otimes T} 
 \end{array}\right]\,
\label{convertpotentials}
\end{eqnarray}
for vectors in color space, we can easily convert the potential operators in
Eqs.~\eqref{pots},~\eqref{Lpu},~\eqref{Oki} from the basis formed by the two
${\bf 3}\otimes{\bf \bar 3}$ color structures $1\otimes \bar 1$ and $T^A \otimes
\bar T^A$ into the color singlet/octet basis used for physical applications. In
this paper only the gauge invariant color singlet configuration of 
the potentials (and the ${\cal O}_{ki}$) will be considered. In order to provide
compact results we will therefore restrict ourselves to the renormalization of
the singlet operators 
\begin{align}
{\cal O}_{k1}^{(s)} =& -\frac{
[{\cal V}_c^{(T)}(\nu)]^2\mu_S^{4\eps}}{m}\, {\rm P_{Singlet}}
 \sum_{\bmp,\bmpp,\bmq} ( f_{0} + f_{1} + 2 f_{2} )\
 \big[ \psi_{\bmpp}^\dagger \,\psi_{\bmp}\,
 \chi_{-\bmpp}^\dagger \, \chi_{-\bmp} \big] \,, \nn\\
{\cal O}_{k2}^{(s)} =& -\frac{
[{\cal V}_c^{(T)}(\nu)]^2\mu_S^{4\eps}}{m}\, {\rm P_{Singlet}}
 \sum_{\bmp,\bmpp,\bmq} ( f_{1} + f_{2} ) \ 
 \big[ \psi_{\bmpp}^\dagger \, \psi_{\bmp}\,
 \chi_{-\bmpp}^\dagger \, \chi_{-\bmp} \big] \,,  \nn\\
{\cal O}_{k3}^{(s)} =& -\frac{
[{\cal V}_c^{(T)}(\nu)]^2\mu_S^{4\eps}}{m}\, {\rm P_{Singlet}}
 \sum_{\bmp,\bmpp,\bmq} (f_0 + f_1) \ 
 \big[ \psi_{\bmpp}^\dagger \, \psi_{\bmp}\,
 \chi_{-\bmpp}^\dagger \, \chi_{-\bmp} \big] \,. \label{OkiSing}
\end{align}
Here ${\rm P_{Singlet}}= ( 1 - 2\, \frac{C_F}{C_A} ) \csI - \frac{2}{C_A} \csT$ is the color singlet projection operator.

All fields, couplings and Wilson coefficients in the above Lagrangian are to be 
understood as bare quantities unless stated otherwise. For
the renormalized quantities, indicated here by the index $R$, we chose the 
usual conventions in $d=4-2\epsilon$ dimensions:
\begin{equation}
\begin{array}{ll}
g_U \,=\, \mu_U^\epsilon\, Z_g \, g_U^R\,,&\quad g_S \,=\, \mu_S^\epsilon\,Z_g \, g_S^R\,,\\
\psi_\bmp = Z^{1/2}_{\psi , \bmp}\; \psi_\bmp^R\,,&\quad Z_{\psi , \bmp} = 1 \!+\! \delta  Z_{\psi , \bmp}\,,\;\;(\psi \to \chi),\\
A^\mu = Z_A^{1/2} \: A_R^\mu\;,&\quad Z_A=1 \!+\! \delta Z_A\,,\\[1 ex]
{\cal V}_i =\mu_S^{2\epsilon}({\cal V}_{i,R} \!+\! \delta {\cal V}_i)\,,&\quad {\cal V}_{ki} =\mu_S^{2\epsilon}({\cal V}_{ki,R} \!+\! \delta {\cal V}_{ki})\,,
\end{array}
\label{RenDef}
\end{equation}
where $\mu_S=m\nu$ and $\mu_U=m\nu^2$ are the soft and ultrasoft
renormalization scales, respectively.
For convenience, we will drop the index $R$ throughout this paper and only
deal with $\msb$ renormalized quantities in the following. We will moreover use the notation
\be
\as \equiv \alpha_s(m \nu) = \frac{g_S^2}{4 \pi} 
\,,\qquad 
\au \equiv \alpha_s(m \nu^2) = \frac{g_U^2}{4 \pi}\,.
\ee

\section{Renormalization Procedure}
\label{sectionrenproc}
 
The aim of this work is to determine the ultrasoft anomalous dimension of
the sum operator coefficients ${\cal V}_{ki}^{(1,T)}(\nu)$ at NLL order. 
We work in Feynman gauge and use the
$\msb$ renormalization scheme in combination with dimensional regularization
($d=4-2\eps$).  
All operators ${\cal O}_{ki}^{(1,T)}$ in Eq.~\eqref{OkiSing} have zero matching 
condition at the hard scale, i.e. \mbox{${\cal
    V}_{ki}^{(1,T)}(\nu=1)=0$}~\cite{HoangStewartultra}. 

Purely ultrasoft loop corrections to potential operators in vNRQCD are
inevitably suppressed by two powers of the velocity $v$.\footnote{
At $\ord(v^0)$
the interaction between ultrasoft $A^0$ gluons and heavy quarks can be removed by a
field redefinition that leaves physical predictions unchanged~\cite{LMR,PhD}.} 
We therefore need at least one additional compensating factor $\sim \as/v$ from
a potential loop in order to obtain the correct $v$-scaling of a
$V_k$ potential~\cite{LMR,ManoharVk}. This is why the 
$V_k$ potential first receives a running at $\ord(\as^2)$ as
indicated by the factor $({\cal V}_c^{(T)})^2$ in the definition of the 
${\cal O}_{ki}^{(1,T)}$. 

\begin{figure}[t]
 \includegraphics[width = 200 pt]{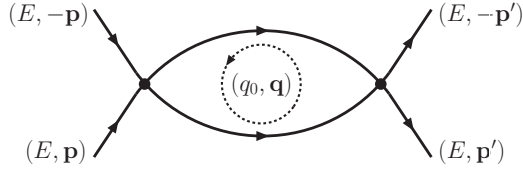}
\caption{External energy/momenta convention for all diagrams with potential loop.
 \label{momenta}}
\end{figure}

Figure~\ref{LLTops} and table~\ref{diagclasses} show relevant UV-divergent
$\ord(\as^2 \au)$ and $\ord(\as^2 \au^2)$ diagrams, which
contribute in Feynman gauge to the anomalous dimension of the $V_k$ potentials at
ultrasoft LL and NLL level, respectively. Each diagram consists of one UV-finite
potential loop and UV-divergent ultrasoft gluon attachments. Many diagrams also
contain ultrasoft derivative operators from higher orders in the multipole expansion. 
They are depicted as nabla ($\nabla$) and Laplace ($\Delta$) operator symbols on heavy 
quark lines or on potential vertices. 
The operator insertions on the lines represent relativistic corrections to the
heavy quark kinetic term $\propto (\partial_0 - \frac{\bmp^2}{2 m})$ (see
Eq.~(\ref{Lus})) due to ultrasoft momentum components flowing through the heavy quark propagators.
The potentials with ultrasoft derivatives are defined in Eq.~\eqref{Lpu} and represent corrections due to ultrasoft momentum exchange between quark and antiquark.
 The corresponding Feynman rules are given in
App.~\ref{feynmanrules}.\footnote{
Such derivative operators also exist for heavy quark lines/vertices which do not carry ultrasoft
momentum components, but then they evaluate to zero. This needs to be taken into
account when showing the equivalence of diagrams that differ by the routing of
ultrasoft momenta. 
} 
In Fig.~\ref{momenta} we show our generic choice
for the external energy and momenta running through the diagrams with a
potential loop. Note that vNRQCD as reviewed in Sec.~\ref{sectiontheory} is
formulated in the center of mass frame. The little arrows on the heavy quark
lines denote positive energy flow and are suppressed in the other graphs
shown in this paper. 

As demonstrated in Ref.~\cite{HoangStewartultra} the additional potential
loop entails the renormalization of the ${\cal O}_{ki}^{(1,T)}$ sum operators
instead of the potentials in Eq.~(\ref{4qVk}). The reason becomes apparent at
the three-loop level, when the potential loop appears together with two  
ultrasoft loops. Some typical example diagrams are shown in
table~\ref{diagclasses}. After the integrations over ultrasoft $\it and$
potential loop momenta their leading divergences typically take the schematic
form~\cite{HoangStewartultra} 
\begin{eqnarray} \label{bad}
  \frac1{|\bmk|}\!\Big(\frac{\mu_S^{2}}{{\bf k}^2}\Big)^\epsilon \bigg[ \frac{1}{2\epsilon^2}
    \Big(\frac{\mu_U^{2}}{E^2}\Big)^{2\epsilon} \!\!
   - \frac{1}{\epsilon^2} \Big(\frac{\mu_U^{2}}{E^2}\Big)^{\epsilon} 
    \bigg] = \frac{1}{|\bmk|}\!\Big[\!- \frac{1}{2\epsilon^2} 
    - \frac{1}{2\epsilon} \ln\!\Big(\frac{\mu_S^2}{{\bf k}^2}\Big) 
    \!+ \ldots\Big] \,,
\end{eqnarray}
where $E$ represents the ultrasoft scale $\sim mv^2$, i.e. the kinetic energy of
the external heavy quarks. 
Inside the square brackets the first term arises from the three-loop diagrams
and the second from two-loop diagrams with one-loop counterterms to remove
subdivergences. 
The factor in front of the square brackets on the left-hand-side comes from the
finite potential loop. The $\ln(\mu_S^2/\bmk^2)$-dependent
$1/\eps$-pole on the right-hand-side of Eq.~\eqref{bad} cannot be
absorbed into counterterms for potentials of the type in Eq.~(\ref{4qVk}), because
this would cause inconsistent anomalous dimensions. This is the reason, why the
sum operators 
${\cal O}_{ki}$ are essential for a consistent renormalization beyond the
one-loop level in vNRQCD. Upon summing the intermediate soft momentum label
$\bmq$, they contain the proper potential prefactors in
Eq.~\eqref{bad} in their operator structure. Therefore the $\delta {\cal
  V}_{ki}$ counterterms only absorb proper, scale independent UV poles and lead
to consistent anomalous dimensions. 

In practice we have to determine the $\delta {\cal V}_{ki}$ from the ultrasoft
UV divergences of two- (LL) and three-loop (NLL) Feynman diagrams 
after the integration over the zero-component of the potential loop momentum
($q_0$), which can be carried out using residues. 
The potential three-momentum ($\bmq$) integration, however, is not carried
out in the renormalization procedure as the $\bmq$-dependence of the remaining
UV-divergences has to be matched onto the label structures of the sum operators
${\cal O}_{ki}$.  
There is a technical subtlety concerning the vNRQCD multipole expansion related
to taking residues with respect to the potential loop momentum $q_0$ in the
upper or the lower complex half plane. Here differences can arise in
the soft momentum label structure in terms of $\bmp$, $\bmpp$ and $\bmq$
resulting from insertions of the ultrasoft derivative operators which are
non-zero only when the corresponding quark line carries an ultrasoft momentum
component. Upon carrying out the label sum the results are unique.
To avoid redundancies in the operator structure of the ${\cal O}_{ki}$ we need
to define a consistent prescription for the respective Feynman graphs in order
to ensure, that the operator basis of the ${\cal O}_{ki}$ required for the
renormalization is unique. 
We evaluate each Feynman diagram with a potential loop considered in this paper
as follows ($q^\mu=(q_0,\bmq)$ being the potential loop momentum): 
\begin{itemize}
\item[1.]
Neglect the kinetic and potential multipole correction operator
insertions for the moment. Pick one internal heavy quark or antiquark line (which carries momentum $\bmq$)
and choose an ultrasoft momentum routing such that this heavy quark/antiquark line does not
carry any ultrasoft momentum. Now, according to the chosen routing insert the required ultrasoft derivative operators in all possible ways.

\item[2.]
Compute the residue of the pole in the complex $q_0$ plane associated with the
propagator without ultrasoft momentum picked in step~1 for the chosen routing and each
vertex/insertion configuration. Multiply with the proper factor
($\pm2\pi i$) according to the residue theorem depending on whether the pole
lies in the upper or lower complex half plane. 

\item[3.]
Perform the ultrasoft loop integrals.

\item[4.]
Repeat steps~1-3 for all internal heavy quark/antiquark lines. 
Sum all expressions obtained in this way and divide the result by two.
 
\end{itemize}
The factor $1/2$ from point~4 compensates for the overcounting, which arises in points~1 and~2 from closing the $q_0$ integration contour in the upper as well as in the lower complex half plane.

From a technical point of view this procedure corresponds to the renormalization of properly defined six-quark operators in order to absorb the ultrasoft UV divergences of the six-leg diagrams, which are generated from cutting the heavy quark/antiquark lines of the potential loop where the residues of step 2 are taken. This is the origin of the notation used in Fig.~\ref{OneloopCuts}.
The two heavy quark fields associated with the cut can then be contracted to form tadpole diagrams with four external legs, which is equivalent to carrying out the label sum for the sum operators.
This procedure yields the same results as using the matrix elements of the sum-operators (see Ref.~\cite{PhD} for details). In Fig.~\ref{OneloopCuts} we visualize the terms arising in point 2 of the above caculational prescription using their analogy to the described six-leg diagrams.

\section{LL Calculation}
\label{sectionLLresults}

\begin{figure}[htp]
 \includegraphics[width = \textwidth]{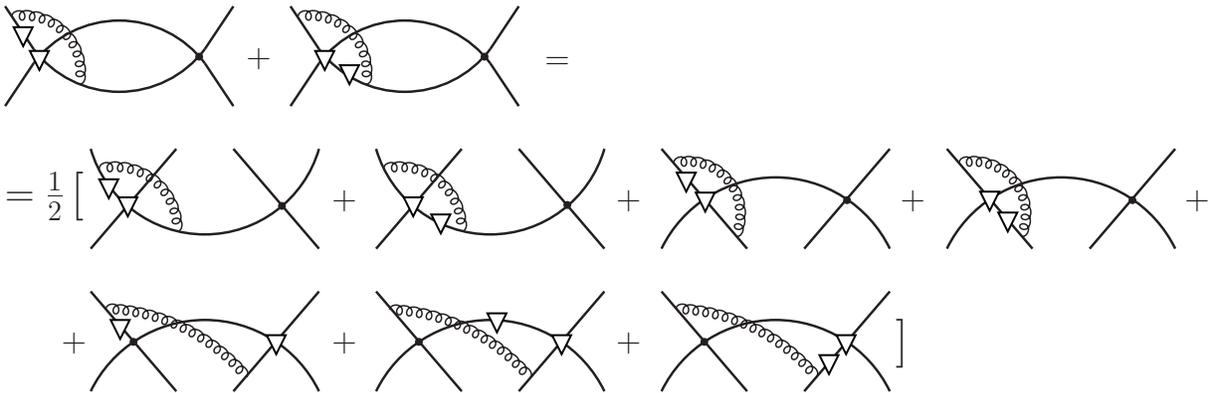}
\caption{Example for a calculation with one ultrasoft loop involving ultrasoft derivative operators. 
 Each of the cut heavy quark/antiquark lines on the right-hand side corresponds to a $q_0$ residue to be taken at this line within the original closed potential loop diagram on the left-hand side according to step 2 of our calculational prescription.
The color structure of the diagrams is understood to be not affected by the cutting, i.e. it is the same on both sides of the equation.
 \label{OneloopCuts}}
\end{figure}

In order to illustrate our computational procedure let us consider the simple
case of the two $\ord(\as^2 \au)$ diagrams shown in Fig.~\ref{OneloopCuts}.
In Feynman gauge they contribute to the LL anomalous dimension of the ${\cal
  O}_{ki}$  and contain ultrasoft derivative operators. The
outcome of steps~1 to 4 are sketched graphically in
Fig.~\ref{OneloopCuts}. Each six-leg diagram
corresponds to picking up the residue of the cut heavy quark/antiquark line as
described in step~2. The overall factor $1/2$ is due to step~4.
Including in addition all possible mirror-graphs and the analogous diagrams with one
$\Delta$-operator, we obtain (up to an overall color factor  not shown) the result
$\frac{- i {\cal V}_c^2 \au }{ \pi  m \epsilon } \int\!\!\! \frac{d^{d-1}q}{(2
  \pi)^{d-1}} \frac{2f_1}{3} + \ord(\epsilon^0)$, 
where in the integrand the term $(f_0+f_1)/4$ arises from the residue in the
upper complex half plane and $(-3f_0+5f_1)/12$ from the residues in the the lower
complex half plane. Upon integration over $\bmq$ both residue contributions
agree.

\begin{figure}[htp]
 \includegraphics[width =0.7 \textwidth]{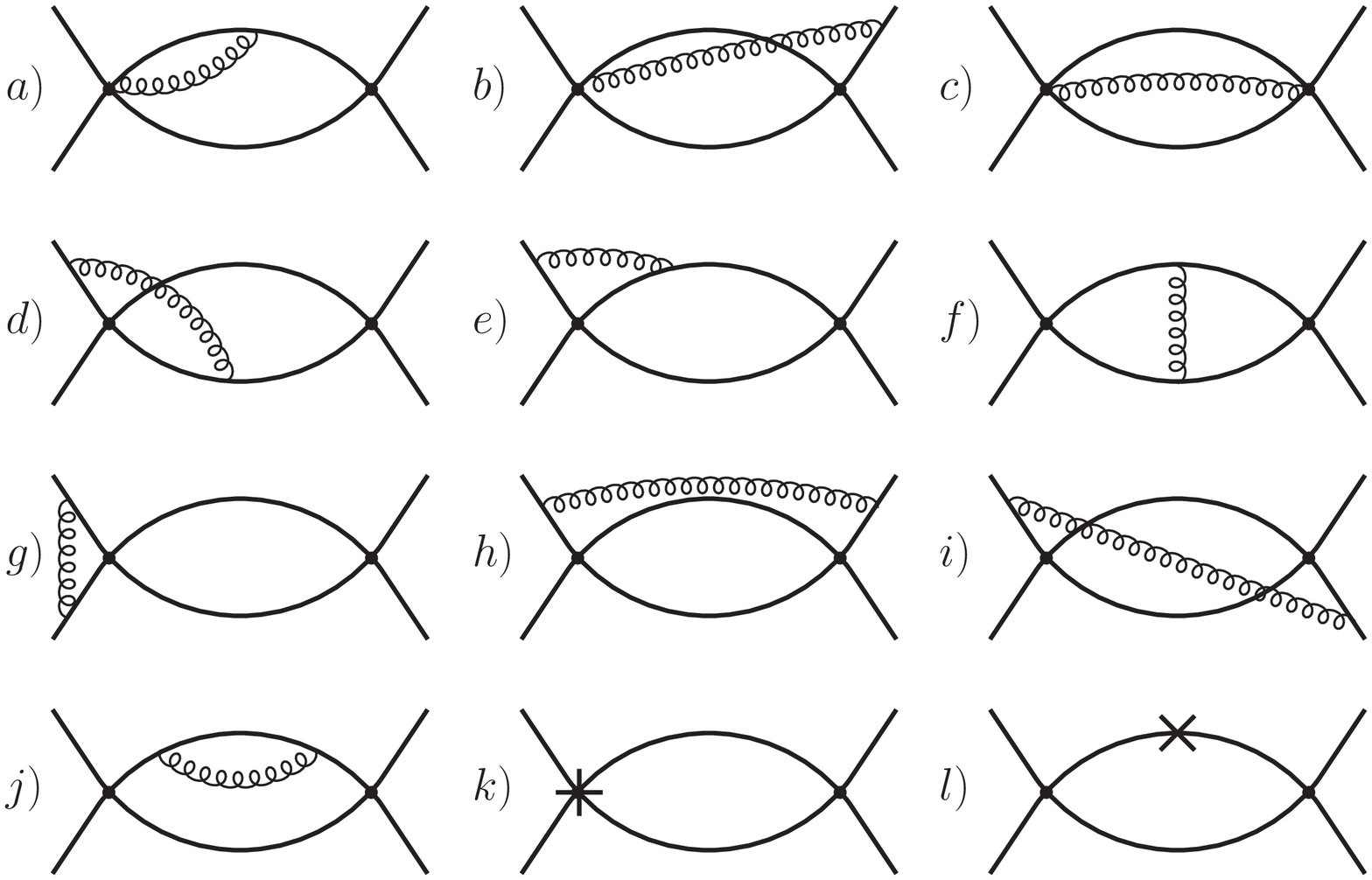}
\caption{Diagram topologies that generate the RG running of the ${\cal O}_{ki}$ at LL level. Up-down and left-right mirror graphs are understood.
 \label{LLTops}}
\end{figure}

We now briefly discuss the full LL ultrasoft renormalization of the 
${\cal O}_{ki}$ in Feynman gauge; for details see Ref.~\cite{PhD}. 
This calculation is also interesting by itself since the available vNRQCD
computations of Refs.~\cite{ManoharVk,HoangStewartultra} were carried out in
Coulomb gauge. Fig.~\ref{LLTops}
shows the relevant diagram topologies. Possible insertions of ultrasoft
derivative operators are not shown.  
\\
The ultrasoft gluons in diagrams~\ref{LLTops}~a-c are spatial ($\bmA$ type)
because there are no four-quark operators with one additional $A^0$ ultrasoft
gluon in vNRQCD. Thus
the vertices in these graphs already contribute two powers of $v$ to the 
velocity counting and we do not need to insert further derivative operators. On
the other hand, diagrams with a spatial $\bmA$ gluon and topologies~d-j 
are either UV finite or their divergence is exactly canceled by one
loop diagrams with $\delta {\cal V}_r$, $\delta {\cal V}_2$ or wavefunction
counterterm insertions. The latter are depicted in diagrams~\ref{LLTops}~k and
l, respectively. The very same applies to the diagrams with an $A^0$ gluon
for topologies d-j, where two kinetic $\nabla$-operators are inserted on
the heavy quark lines within the ultrasoft loop. 

The remaining diagrams~d-j with an $A^0$ gluon having the correct $v$ scaling
contain either one $\Delta$- or two $\nabla$-operator insertions, of
which at least one is  the $\nabla$-potential in Eq.~(\ref{NablaPot}). Adding
them to the $\bmA$ diagrams~\ref{LLTops}~a-c gives  
\begin{align}
 -& i C_A \lefteqn{  \frac{\au  ({\cal V}_c^{(T)})^2 \mu_S^{4 \epsilon}}{3 m   \pi  \epsilon } } \nn\\
&
\times \int\!\!\! \frac{d^{d-1}q}{(2 \pi)^{d-1}}
\Big[\, C_F  (C_A\!-\!2 C_F) (f_0\!+\!f_1\!+\!2 f_2) \csI \,+\,
 (C_A\!-\!4 C_F)(f_1\!+\!f_2) \csT \,\Big].\label{LLresult}
\end{align}
The result is exactly the same as obtained in
Refs.~\cite{ManoharVk,HoangStewartultra} where Coulomb gauge has been used. We
note that the Coulomb gauge computation at LL order is substantially simpler
since in Coulomb gauge $A^0$ gluons cannot propagate and the number of diagrams
that have to be considered is reduced considerably. However, the Coulomb gauge
calculation becomes a nightmare at the two-loop level and seems entirely
unfeasible.  We also note that we
regularize IR divergences by adopting an off-shell configuration for the
external heavy quark lines with $E\neq \bmp^2/2m = (\bmp^\prime)^2/2m$.
 
From Eq.~\eqref{LLresult} we can easily read off the already known $\ord(\au)$
color singlet counterterms 
\begin{align}
\delta {\cal V}_{k1}^{(s)}=- C_A C_F (C_A-2C_F) \frac{\au}{3 \pi \eps} \,; \qquad \delta {\cal V}_{k2}^{(s)}= C_A C_F (C_A-4C_F) \frac{\au}{3 \pi \eps}\,.
\label{OneLoopdeltaVki}
\end{align}

\section{NLL Calculation}
\label{sectionCalc}

Let us now consider the divergent $\ord(\as^2\, \au^2)$ diagrams contributing to
the ultrasoft NLL anomalous dimension of the ${\cal V}_{ki}$, i.e. diagrams with
one potential and two ultrasoft loops. In table~\ref{diagclasses} we have
classified the relevant diagrams without gluon selfenergy with respect to their
vertices and kinematic 
operator insertions. 
For each class one sample
diagram with a specific configuration of vertices and
insertions is shown graphically.
Diagrams with all possible permutations of the vertices and different
attachments of the gluons to the heavy quark lines are understood to belong to
each class.  The respective contribution in the last column of
table~\ref{diagclasses} denotes the sum of the divergent parts of all class
members. Contribution I covers classes 1 and 2, contribution iv covers classes 12 and 13.
In the following we will call the diagram classes 1-8 ``Abelian'' and 9-14 ``non-Abelian''. The relevant
Feynman gauge Feynman rules are found in Appendix~\ref{feynmanrules}.\\

Besides the Abelian and non-Abelian diagrams in table~\ref{diagclasses} there are
$\ord(\as^2\, \au^2)$ diagrams containing one-loop gluon selfenergy subdiagrams
with ultrasoft light fermion, ghost and gluon bubbles. For the
calculation of these diagrams we take the graphs with only one ultrasoft loop in
Fig.~\ref{LLTops}
and replace the ultrasoft gluon propagators by their one-loop correction. 
Note that in Feynman gauge the selfenergy insertion is a non-diagonal matrix in
the components of the ultrasoft gluon field and can couple 
the zero-component ($A^0$) to the spatial components ($\bmA$) of the ultrasoft
gluon field due to its relativistic nature.  
Therefore we have to take into account diagrams with one $A^0$-vertex, one $\bmA$-vertex and
 a single $\nabla$-operator insertion, in addition to the diagrams we obtain by simply
inserting the selfenergy bubble into the $\ord(\as^2 \au)$ diagrams of the
previous section. 
After subtraction of the selfenergy subdivergences by adequate counterdiagrams
the sum of all ultrasoft UV divergences from the three-loop $\ord(\as^2\,
\au^2)$ diagrams with a gluon selfenergy, reads:
\begin{eqnarray}
\lefteqn{
 \bigg[ \frac{C_A (5 C_A-4 T n_f)}{72 \epsilon^2}-\frac{C_A (31 C_A-20 T n_f)}{216 \epsilon} \bigg] \frac{i  \au^2 ({\cal V}_c^{(T)})^2 \mu_S^{4 \epsilon}}{m\, \pi^2}}&&
\nn\\[1 ex]
&&
\times \int\!\!\! \frac{d^{d-1}q}{(2 \pi)^{d-1}}
\Big[\, C_F  (C_A\!-\!2 C_F) (f_0\!+\!f_1\!+\!2 f_2) \csI \,+\,
 (C_A\!-\!4 C_F)(f_1\!+\!f_2) \csT \,\Big],
\label{VkSelfRes}
\end{eqnarray}
where $T=\frac12$ in the fundamental representation of $SU(3)$ and $n_f$ is the number of light quark flavors.
The result contributes to the counterterms $\delta {\cal V}_{k1}^{(1)}$ and $\delta {\cal V}_{k2}^{(T)}$. The gauge invariant contributions from light fermion loops $\propto n_f$ in Eq.~\eqref{VkSelfRes} were also determined in Coulomb gauge~\cite{StahlhofenTUM}.\\

\begin{table}[htp]
\begin{tabular}{|c|c|c|c|c|c|c|}
\hline
 &&\multicolumn{4}{c|}{vertices}& \\ \cline{3-6}
\raisebox{2 ex}{class} & \raisebox{2 ex}{example diagram} & potential & quark-gluon& 3 gluon& kin. ins. & \raisebox{2 ex}{contrib.}\\
\hline
\hline
& & $1\times V_c$ & $1\times {\bf A}$ &&& \\
\raisebox{2 ex}{1}& \begin{picture}(80,0)(0,5) \includegraphics[width = 80pt]{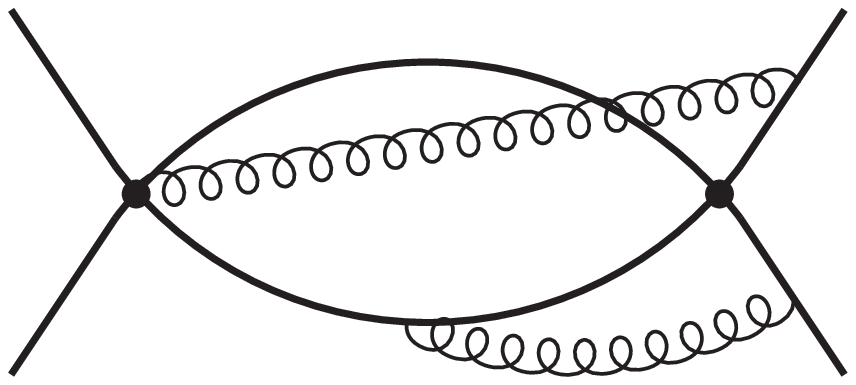} \end{picture} &
 $1\times {\bf A}\!\cdot\! {\bf k}$  &  $2 \times A^0$  & \raisebox{2 ex}{-} &\raisebox{2 ex}{-}&  \\ 
\cline{1-6}
&& &  & & & \raisebox{2 ex}[-2 ex]{I} \\
\raisebox{2 ex}{2}& \begin{picture}(80,0)(0,5) \includegraphics[width = 80pt]{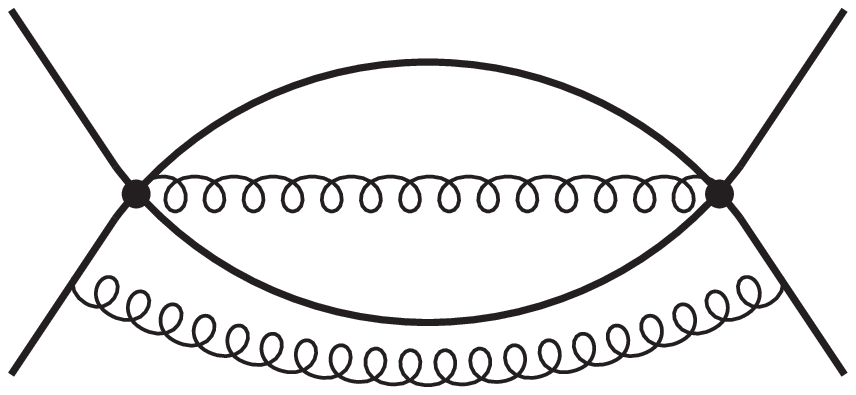} \end{picture} & 
 \raisebox{2 ex}{$2\times {\bf A}\!\cdot\! {\bf k}$}  &\raisebox{2 ex}{$2 \times A^0$}  &\raisebox{2 ex}{-}&\raisebox{2 ex}{-}& \\
\hline
&&  & $2\times {\bf A}$   & & &  \\
\raisebox{2 ex}{3}& \begin{picture}(80,0)(0,4) \includegraphics[width = 80pt]{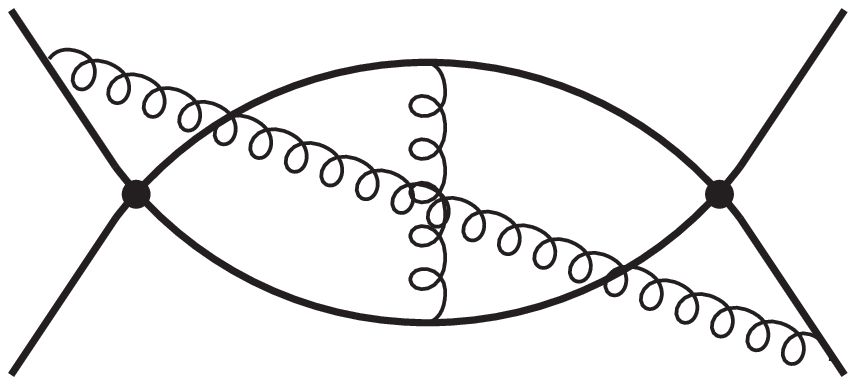} \end{picture} &
  \raisebox{2 ex}{$2\times V_c$}   &  $2 \times A^0$  &\raisebox{2 ex}{-}&\raisebox{2 ex}{-}& \raisebox{2 ex}{0} \\
\hline
&&  &  & & &  \\
\raisebox{2 ex}{4}& \begin{picture}(80,0)(0,3) \includegraphics[width = 80pt]{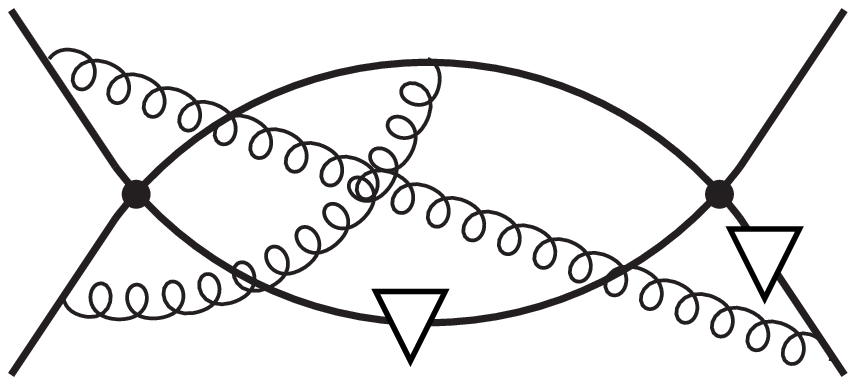} \end{picture} &
 \raisebox{2 ex}{$2\times V_c$}  & \raisebox{2 ex}{$4 \times A^0$} &\raisebox{2 ex}{-}& \raisebox{2 ex}{$2\times \nabla\!\cdot\!\bmp$} & \raisebox{2 ex}{0} \\
\hline
&&  &  & & &  \\
\raisebox{2 ex}{5}& \begin{picture}(80,0)(0,4) \includegraphics[width = 80pt]{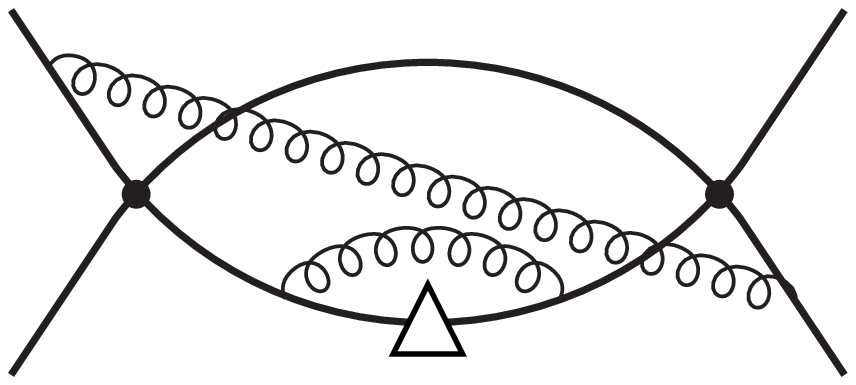} \end{picture} &
 \raisebox{2 ex}{$2\times V_c$} & \raisebox{2 ex}{$4 \times A^0$} &\raisebox{2 ex}{-}& \raisebox{2 ex}{$1\times \nabla^2$} & \raisebox{2 ex}{II} \\
\hline
&&  &  & & &  \\
\raisebox{2 ex}{6}& \begin{picture}(80,0)(0,4) \includegraphics[width = 80pt]{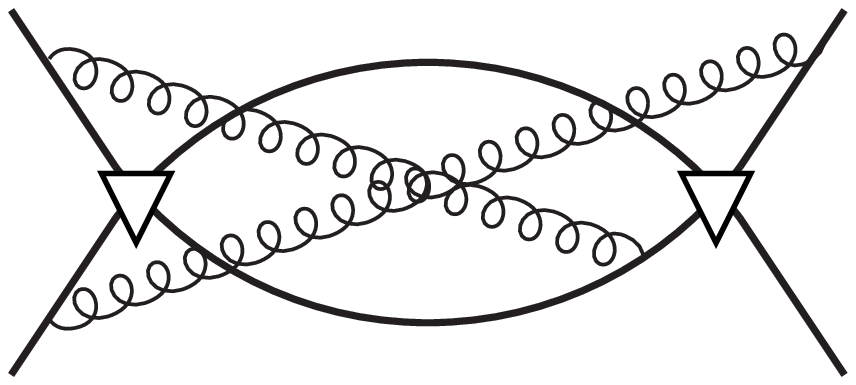} \end{picture} &
  \raisebox{2 ex}{$2\times \nabla\!\cdot\! {\bf k}$}  &  \raisebox{2 ex}{$4 \times A^0$} &\raisebox{2 ex}{-}&\raisebox{2 ex}{-}& \raisebox{2 ex}{III} \\
\hline
&& $1\times V_c$   &  & & &  \\
\raisebox{2 ex}{7}& \begin{picture}(80,0)(0,4) \includegraphics[width = 80pt]{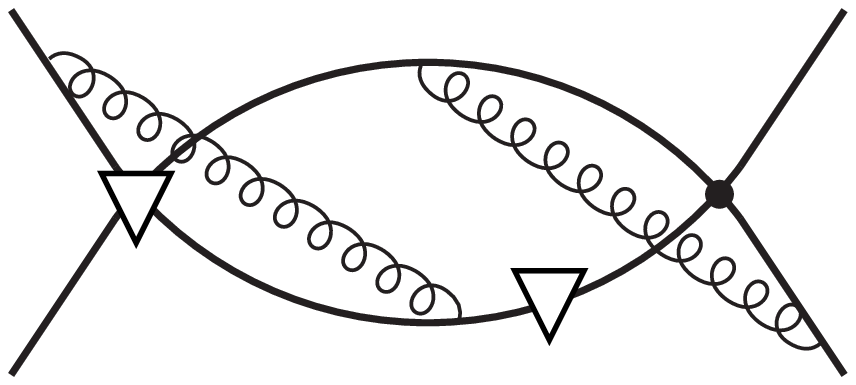} \end{picture} &
 $1\times \nabla\!\cdot\! {\bf k}$  &  \raisebox{2 ex}{$4 \times A^0$} &\raisebox{2 ex}{-}&\raisebox{2 ex}{$1\times \nabla\!\cdot\!\bmp$}& \raisebox{2 ex}{IV} \\
\hline
&&  &  & & &  \\
\raisebox{2 ex}{8}& \begin{picture}(80,0)(0,4) \includegraphics[width = 80pt]{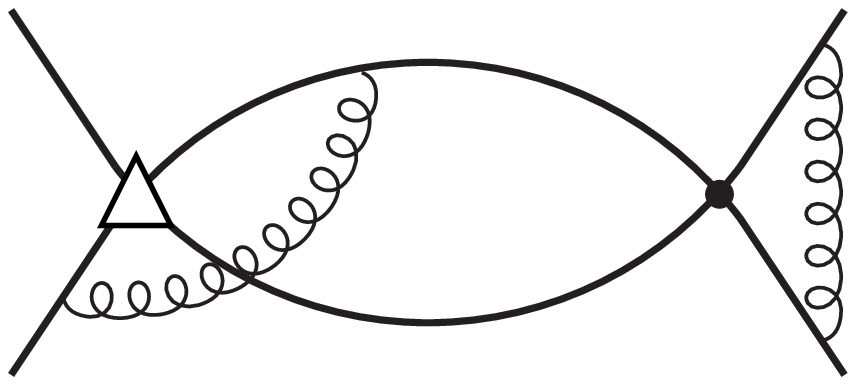} \end{picture} &
  \raisebox{2 ex}{$1\times \nabla^2$}  &  \raisebox{2 ex}{$4 \times A^0$} &\raisebox{2 ex}{-}&\raisebox{2 ex}{-}& \raisebox{2 ex}{V} \\
\hline
&&  & $2\times {\bf A}$   & & &  \\
\raisebox{2 ex}{9}& \begin{picture}(80,0)(0,4) \includegraphics[width = 80pt]{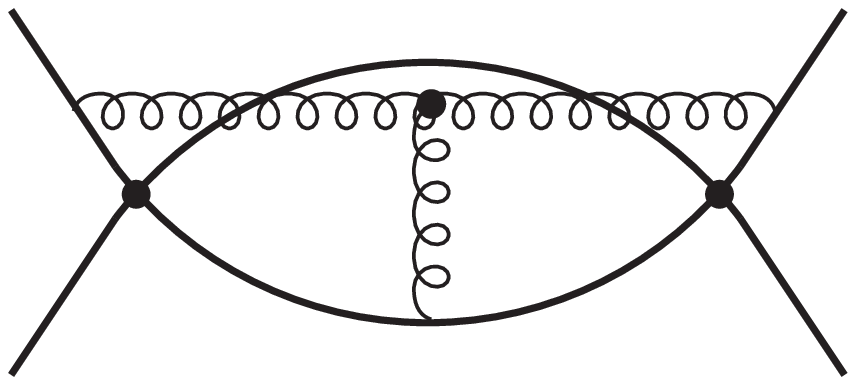} \end{picture} &
  \raisebox{2 ex}{$2\times V_c$}   &  $1 \times A^0$  &\raisebox{2 ex}{$\bmA^2 A^0$}&\raisebox{2 ex}{-}& \raisebox{2 ex}{i} \\
\hline
&&  & $1\times {\bf A}$  & & &  \\
\raisebox{2 ex}{10}& \begin{picture}(80,0)(0,3) \includegraphics[width = 80pt]{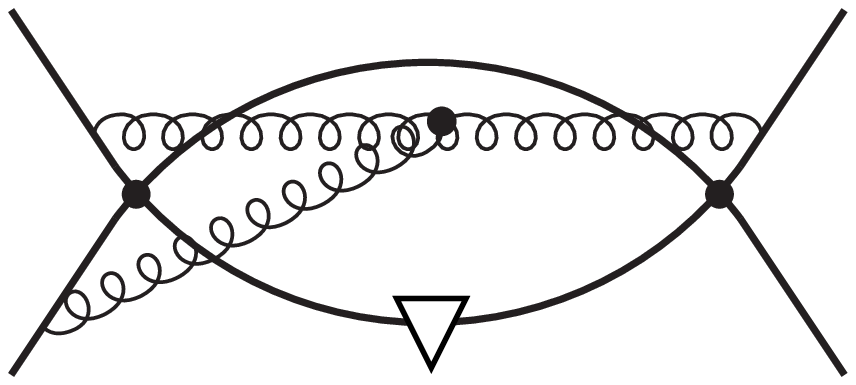} \end{picture} &
 \raisebox{2 ex}{$2\times V_c$}  & {$2 \times A^0$} &\raisebox{2 ex}{$\bmA (A^0)^2$}& \raisebox{2 ex}{$1\times \nabla\!\cdot\!\bmp$} & \raisebox{2 ex}{ii} \\
\hline
&& $1\times V_c$     & $1\times {\bf A}$  & & &  \\
\raisebox{2 ex}{11}& \begin{picture}(80,0)(0,4) \includegraphics[width = 80pt]{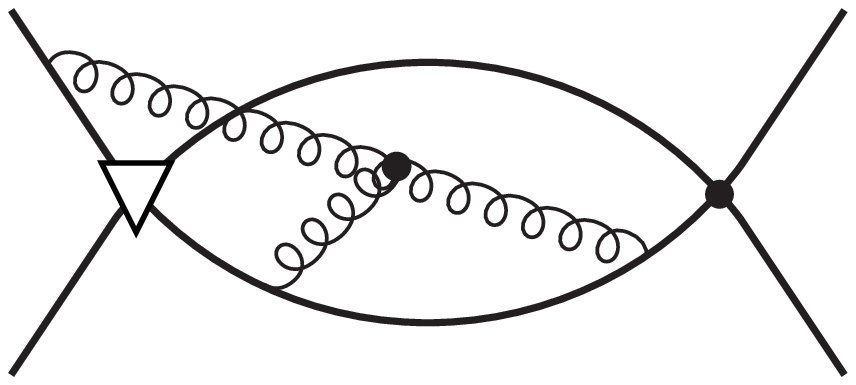} \end{picture} &
$1\times \nabla\!\cdot\! {\bf k}$  & $2 \times A^0$ &\raisebox{2 ex}{$\bmA (A^0)^2$}&\raisebox{2 ex}{-}& \raisebox{2 ex}{iii} \\
\hline
& & $1\times V_c$ & $1\times {\bf A}$ &&& \\
\raisebox{2 ex}{12}& \begin{picture}(80,0)(0,4) \includegraphics[width = 80pt]{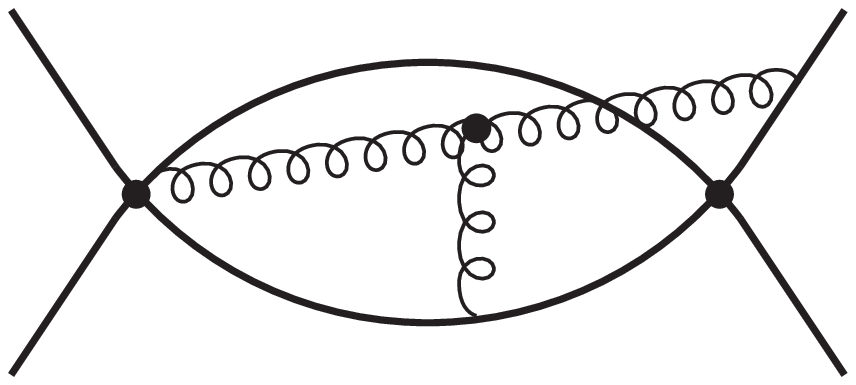} \end{picture} &
 $1\times {\bf A}\!\cdot\! {\bf k}$  &  $1 \times A^0$  &\raisebox{2 ex}{$\bmA^2 A^0$}&\raisebox{2 ex}{-}&\\ 
\cline{1-6}
&& &  & & & \raisebox{2 ex}[-2 ex]{iv} \\
\raisebox{2 ex}{13}& \begin{picture}(80,0)(0,4) \includegraphics[width = 80pt]{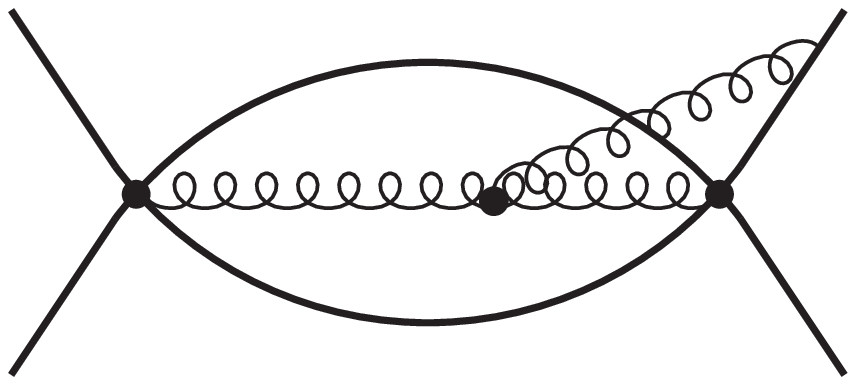} \end{picture} &
 \raisebox{2 ex}{$ 2\times {\bf A}\!\cdot\! {\bf k}$  }&\raisebox{2 ex}{  $1 \times A^0$  }&\raisebox{2 ex}{$\bmA^2 A^0$}&\raisebox{2 ex}{-}&  \\ 
\hline
& & $1\times V_c$ & &&& \\
\raisebox{2 ex}{14}& \begin{picture}(80,0)(0,4) \includegraphics[width = 80pt]{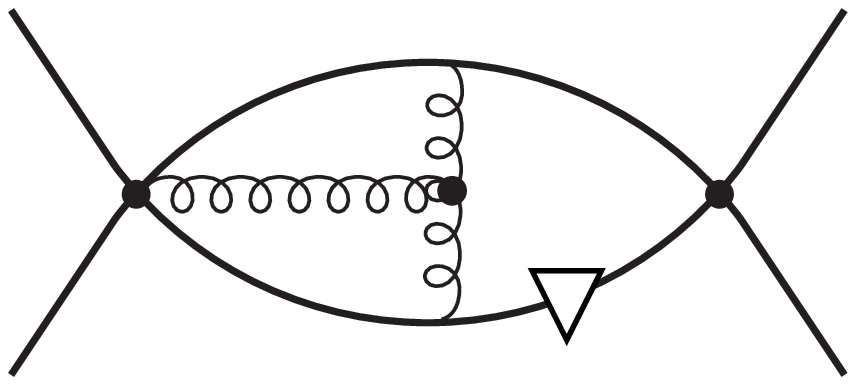} \end{picture} &
 $1\times {\bf A}\!\cdot\! {\bf k}$  &\raisebox{2 ex}{$2 \times A^0$} & \raisebox{2 ex}{$\bmA (A^0)^2$} & \raisebox{2 ex}{$1\times \nabla\!\cdot\!\bmp$}&\raisebox{2 ex}{v}  \\ 
\hline
& & $1\times \nabla\!\cdot\! {\bf k}$ &  &&& \\
\raisebox{2 ex}{15}& \begin{picture}(80,0)(0,4) \includegraphics[width = 80pt]{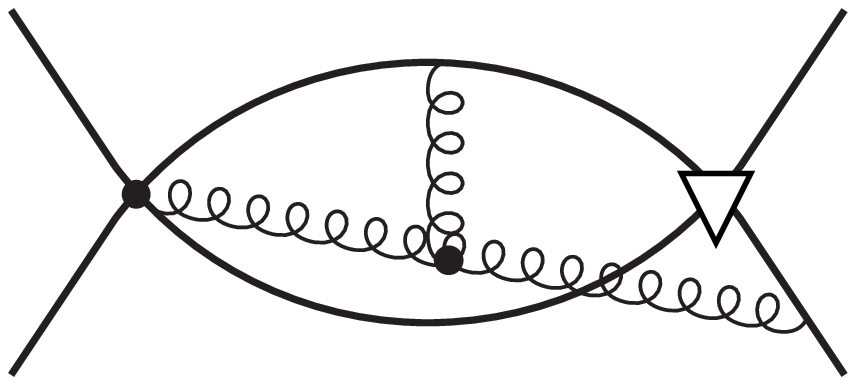} \end{picture} &
 $1\times {\bf A}\!\cdot\! {\bf k}$  &\raisebox{2 ex}{$2 \times A^0$} & \raisebox{2 ex}{$\bmA (A^0)^2$} &\raisebox{2 ex}{-}&\raisebox{2 ex}{vi}  \\ 
\hline

\end{tabular}
\caption{Diagram classes with one potential and two ultrasoft loops (without gluon self energy). 
Diagrams with all possible permutations of the vertices and different attachments of the gluons to the heavy quark lines are understood to belong to each class.
The vNRQCD Feynman rules for the respective vertices, propagators and kinematic operator insertions are given in App.~\ref{feynmanrules}. \label{diagclasses}} 
\end{table}

In table~\ref{AbNonAbRes} we present results for the UV divergent parts of the
Abelian and non-Abelian diagrams classified in table~\ref{diagclasses} after all
one- and two-loop subdivergences that contribute to the renormalization of the
$1/m^2$~potentials~\cite{amis,V2Vr} have been subtracted.\footnote{The results in table~\ref{AbNonAbRes} can also be found in Eqs.~(6.33) and~(6.34) (Abelian contributions without wavefunction renormalization contributions) and table~6.9 (non-Abelian contributions) of Ref.~\cite{PhD}. In table~6.8 of Ref.~\cite{PhD} a result for the non-Abelian contributions is shown that is based on a calculation of six-leg diagrams, as explained at the end of Sec.~\ref{sectionrenproc}, using an off-shell configuration for the two lines coming from cutting the intermediate heavy quark propagator. This off-shell configuration disagrees with the on-shell condition arising from doing the potential energy integration by the residue theorem, and the result is not relevant for the computation intended in this work.
}
The net contributions from the Abelian diagram classes 3 and 4 vanish, because the
diagrams are either finite or the UV-divergences are cancelled exactly by the
respective diagrams with $1/m^2$~potential counterterms. 
\begin{table}[ht]
\begin{center}
\begin{tabular}{|c|l|}
\hline
contribution &\rule[-2.2 ex]{0 ex}{6 ex} \hspace{25ex} result $\cdot \Big(\frac{i\, {\cal V}_c^2 \au^2}{m} \Big)^{\!\!-1}$ \\
\hline
I &\rule[-2.5 ex]{0 ex}{7 ex} $ \big( \frac{1}{8\, \epsilon^2}-\frac{1}{4\, \epsilon } \big) \frac{C_A^2}{\pi^2}
 \scriptstyle \big[ C_F(C_A \!-\!2 C_F)(f_0\!+\!f_1\!+\!2 f_2) \csI \,+\, (C_A\!-\!4 C_F) (f_1\!+\!f_2)\, \csT  \big]$ \\

\hline
II+III+IV+V &\rule[-2.5 ex]{0 ex}{6 ex} 
$ \big(\!-\frac{1}{24\, \epsilon^2} + \frac{1}{8\, \epsilon}\, \big) \frac{C_A^2}{\pi^2}
\scriptstyle \big[ C_F(C_A \!-\!2 C_F)(f_0\!+\!f_1\!+\!2 f_2) \csI \,+\, (C_A\!-\!4 C_F) (f_1\!+\!f_2)\, \csT  \big]$ \\

\hline
i+ii &\rule[-2.2 ex]{0 ex}{6 ex}$ 
\frac{ \bmq^2 }{48\, \epsilon \, (\bmq-\bmp)^2 (\bmpp-\bmq)^2 (2 m E-\bmq^2)}\, \scriptstyle C_A \big[C_F (C_A^2 -6 C_A C_F + 8 C_F^2)\csI  + (C_A^2 -6 C_F C_A +12 C_F^2)\csT \big]$ \\

\hline
    &\rule[-2.2 ex]{0 ex}{6 ex}$\frac{C_A}{16 \, \pi ^2} \Big[
       -\frac1{\epsilon^2} \scriptstyle \,C_A \, \big[ \, (C_A-2 C_F) C_F (f_0+f_1) \csI+(C_A-4 C_F) f_1 \csT \, \big] \, + $ \\
iii &\rule[-2.2 ex]{0 ex}{6 ex}$ {\scriptstyle + \,} \frac1{9 \epsilon} \scriptstyle \big[ C_F(C_A-2 C_F)\csI \big\{ f_0 [6 C_A+(2 C_A-12 C_F) \pi ^2]
        -2 f_1 [-3 C_A+(2 C_A-6 C_F) \pi ^2] \big\}+ $\\ 
    &\rule[-2.2 ex]{0 ex}{6 ex}$\scriptstyle +\, \csT \big\{ 12 C_F(2 C_F-C_A)\pi^2 f_0   +f_1 [6 C_A (C_A-4 C_F)-4 (C_A^2-4 C_F C_A+6 C_F^2) \pi ^2]            
      \big\}\big] \Big] $
 \\
\hline
        &\rule[-2.2 ex]{0 ex}{6 ex} $\frac{C_A}{16 \, \pi ^2}  \Big[ \frac1{\epsilon^2} \scriptstyle \,C_A \, \big[ \, (C_A-2 C_F) C_F
             (f_0+f_1) \csI+(C_A-4 C_F) f_1 \csT \, \big] \, + $ \\
        &\rule[-2.2 ex]{0 ex}{6 ex}${\scriptstyle+}\frac1{9 \epsilon } \scriptstyle \big[ C_F(C_A-2 C_F) \csI \big\{-5 f_0
          [6 C_A+(2 C_A-12 C_F) \pi ^2]+10 f_1 [-3 C_A+(2 C_A-6 C_F) \pi^2] \scriptstyle  -4 C_A f_2 [12-2\pi^2] \big\} $\\

\raisebox{3 ex}[3 ex]{iv+v+vi} &\rule[-2.2 ex]{0 ex}{6 ex}$\scriptstyle + \csT \big\{ 60 (C_A-2 C_F) C_F \pi ^2 f_0 + f_1 [-30 C_A (C_A-4 C_F)+20 (C_A^2-4 C_F C_A+6 C_F^2) \pi ^2]+$\\
        &\rule[-2.2 ex]{0 ex}{6 ex}\hspace{7ex} $\scriptstyle +8 C_A f_2 [-3 C_A+12 C_F+(2C_A-2C_F) \pi ^2] \big\}  \big] \Big]  $\\
\hline
\end{tabular}
\caption{UV divergent contributions before the $\bmq$ integration from
  three-loop diagrams with one potential and two ultrasoft loops as defined in
  table~\ref{diagclasses}. Subdivergences have been subtracted by one- and
  two-loop counterdiagrams. The soft three-momentum structure is (mainly)
  encoded in the $f_i$ functions of Eq.~\eqref{ffunctions}.
\label{AbNonAbRes}
}
\end{center}
\end{table}
The results for I and II+III+IV+V also contain contributions from external
wavefunction diagrams, i.e. heavy quark selfenergies on the four external
legs. Two-loop heavy quark selfenergies with a triple gluon vertex vanish. 

In order to avoid singularities of infrared origin we again assigned a uniform
off-shellness to each of the four external legs. After the subtraction of
subdivergences the logarithmic dependence on the off-shellness vanishes.
 
Note that apart from the external off-shellness
the ultrasoft loops in the six-leg diagrams also involve the physical
off-shellness of the intermediate  
(uncut) heavy quark line, $2(E-\frac{\bmq^2}{2m})$. So the nontrivial ultrasoft 
two-loop diagrams that have to be calculated represent complicated
multi-scale integrals.

We solved the integrals using partial fraction decomposition and integration by
parts techniques. To handle the larger number of diagrams (corresponding to 
$\ord(10^4)$ six-leg diagrams) we developed {\tt Mathematica} codes for the
generation and the evaluation of the  
corresponding amplitudes. For the $\eps=\frac{4-D}{2}$ expansion of the
hypergeometric functions resulting from the two-loop integrations we used the
{\tt Mathematica} package {\tt HypExp~1.1}~\cite{HypExp} and checked the
results numerically. In order to automatize the determination of the color
structures of the different three-loop topologies we moreover employed a few
routines 
from the {\tt FeynCalc~4.1} 
package~\cite{FeynCalc}.

From the $\ord(\au^2)$ UV divergences in table~\ref{AbNonAbRes} and
Eq.~\eqref{VkSelfRes} we can now determine the sum-operator counterterms $\delta
{\cal V}_{ki}$. They take the generic form 
\begin{eqnarray}
\delta {\cal V}_{ki} &=& A_i\, \frac{\au-\as}{\epsilon} \;+\; {\tilde A_i}\, \frac{\au^2-\as^2}{\epsilon^2} \;+\; B_i\, \frac{\au^2-\as^2}{\epsilon}\;.
\label{GendeltaVki}
\end{eqnarray}
In addition to Eq.~\eqref{OneLoopdeltaVki} we included here, following
Refs.~\cite{HoangStewartultra} and~\cite{zerobin}, so-called soft pull-up terms
at $\ord(\as)$ and $\ord(\as^2)$. They originate from the zerobin subtractions
in corresponding amplitudes with one and two soft loops~\cite{zerobin}. 
The ultrasoft one-loop coefficients $A_i$ were given in
Ref.~\cite{HoangStewartultra} and can be read off
Eq.~\eqref{OneLoopdeltaVki}. They read in the color singlet channel
\begin{eqnarray}
A_{1}^{(s)} \!&=&\! -\frac{C_A C_F (C_A\!-\!2C_F)}{3\pi}\,,\qquad  A_{2}^{(s)} \;=\; \frac{C_A C_F(C_A\!-\!4C_F)}{3\pi}\,, \qquad A_{3}^{(s)}=0\,.
\label{Ai}
\end{eqnarray}
For the anomalous dimension of the ${\cal V}_{ki}(\nu)$ to be finite, the
relation 
\begin{eqnarray}
{\tilde A_i} \,=\, - \frac{\beta_0}{8 \pi}\,A_i 
\label{epscheck}
\end{eqnarray}
is a necessary condition.

Summing all Abelian and non-Abelian contributions in table~\ref{AbNonAbRes} except for contribution (i+ii), which we will discuss later, and projecting the result on the physical color singlet channel we find:
\begin{align}
{\rm I+II+III+IV+V+iii+iv+v+vi} \;=\; C_A^2 \bigg[ \frac{1}{12 \eps^2} -\frac{7}{24 \eps} -\frac{\pi^2}{18 \eps} \bigg] \frac{i  \au^2 ({\cal V}_c^{(T)})^2 \mu_S^{4 \epsilon}}{m\, \pi^2} \; \nn\\
\times \int\!\!\! \frac{d^{d-1}q}{(2 \pi)^{d-1}}
\Big[\, C_F  (C_A\!-\!2 C_F) (f_0\!+\!f_1\!+\!2 f_2) -
 C_F (C_A\!-\!4 C_F)(f_1\!+\!f_2)  \Big] {\rm P_{Singlet}}\,.
\label{AbNonAbResSing}
\end{align}
The ${\cal O}(\alpha_U^2/\epsilon^2)$ terms of Eq.~\eqref{VkSelfRes} and Eq.~\eqref{AbNonAbResSing}
 satisfy Eq.~(\ref{epscheck}), which provides an important cross-check of
our calculation. From the respective ${\cal O}(\alpha_U^2/\epsilon)$ terms we determine the coefficients
\begin{align}
B_1^{(s)} &=-\frac{C_A (C_A-2 C_F) C_F \big[C_A(47+6\pi^2) - 10\, n_f T \big]}{108 \pi^2} \,, \nn\\
B_2^{(s)} &= \frac{C_A (C_A-4 C_F) C_F \big[C_A(47+6\pi^2) - 10\, n_f T \big]}{108 \pi^2}
\,.
\label{Bi}
\end{align}

\begin{figure}[t]
 \includegraphics[width = 120 pt]{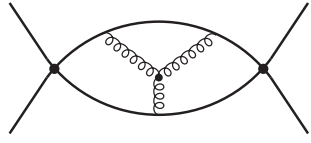}
\caption{
Mercedes star type diagram contained in diagram class 9 (in table~\ref{diagclasses}). Together with the Mercedes star diagrams of class 10 it gives rise to the non-Abelian contribution (i+ii) in table~\ref{AbNonAbRes}. The upside down graph is understood.
 \label{mercedes}}
\end{figure}
The contribution (i+ii) in table~\ref{AbNonAbRes} is generated exclusively by those diagrams of class 9 and 10 which belong to the subclass with the Mercedes star topology shown in Fig.~\ref{mercedes}. All other diagrams of class 9 and 10 add up to zero. The result requires a separate discussion.
It has already been computed in Ref.~\cite{PhD}, but not considered there for numerical examinations. 
Although its effects are numerically tiny, they are interesting conceptually and worth to explain in 
some detail.
Up to a color factor the term (i+ii) has the form 
\begin{align}
\frac{\bmq^2}{48 \epsilon \, (\bmq-\bmp)^2 (\bmpp-\bmq)^2 (2 m E-\bmq^2)}\,.
\label{iplusiiform}
\end{align}
If one carries out the renormalization procedure based on 5-loop current-current
correlators instead of quark-antiquark scattering amplitudes, as realized in
Ref.~\cite{3loop} for the computation of the 3-loop NNLL non-mixing
contributions to the anomalous dimension of $c_1$, this term generates additional
ultrasoft NLL terms in the counterterms of the $1/m^2$~potential coefficient
${\cal V}_r$ and of the coefficient of the $1/m^2$-suppressed S-wave 
current $c_2$.
\begin{figure}[t]
 \includegraphics[width = 0.55\textwidth]{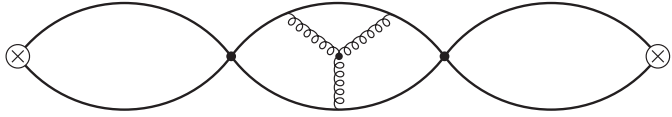}
\caption{
5-loop current-current correlator with a Mercedes star subdiagram. It is relevant in the current-current correlator based renormalization procedure associated with the divergent term (i+ii) in table~\ref{AbNonAbRes}. In this approach one also has to take into account the 5-loop diagrams, where the Mercedes star is inserted into the left and right potential loop. The latter (partly) contribute to the renormalization of the $1/m^2$-suppressed production/annihilation currents.
 \label{5loop}}
\end{figure}
These contributions have not been considered in the literature before. 
They were also not contained in the results of Ref.~\cite{V2Vr} since they do not arise for scattering diagrams with equal external off-shellness, which was used for regulating IR divergences in Ref.~\cite{V2Vr}.
Figure~\ref{5loop} shows a sketch of one of the 5-loop current-current correlator diagrams involved in the renormalization procedure following Ref.~\cite{3loop}. The divergences of these 5-loop diagrams have to be canceled by 3-loop current-current diagams without ultrasoft gluons, but with one of the potentials/currents replaced by the respective $1/m^2$ suppressed counterterm. This approach yields an additional contribution to the counterterm $\delta {\cal V}_r^{(s)}$ to be added to the expression given in Ref.~\cite{V2Vr}. It results from insertions corresponding to the term (i+ii) in table~\ref{AbNonAbRes} into a current-current correlator and reads 
\begin{align}
\Delta \delta {\cal V}_r^{(s)} = 
2 (4 \pi \as)\, \Delta B_\bmp^{(s)}\,;\qquad \Delta B_\bmp^{(s)}  = \frac{C_A C_F^2}{24} \,. 
\label{Bpshift}
\end{align}

Alternatively one might write Eq.~\eqref{iplusiiform} as
\begin{align}
-\frac{f_0+f_1}{96\, \epsilon} + 
\frac{2mE}{48\, \epsilon \, (\bmq-\bmp)^2 (\bmpp-\bmq)^2 (2 m E - \bmq^2)}
\label{split}
\end{align}
and directly absorb the ultrasoft divergence $\propto f_0+f_1$ into the counterterms $\delta{\cal V}_{k3}$, while the counterterm $\delta {\cal V}_r^{(s)}$ of Ref.~\cite{V2Vr} and all other results in that reference remain unchanged.
This leads to the coefficient
\begin{align}
B_3^{(s)} & = \frac{C_A C_F^3}{24} \,.
\label{B3}
\end{align}
The second term of Eq.~\eqref{split} cannot contribute to the
anomalous dimension of $c_1$, because the integrations over the soft
momenta $\bmq$, $\bmp$, $\bmpp$ give finite results, when the
structure is inserted into a current-current correlator. It would rather
contribute to the running of $c_2$, which becomes clear by rewriting its
numerator as $2 m E = 1/2( \bmp^2 + \bmpp^2)$ plus scaleless integrals (over
$\bmp$ and $\bmpp$). The remaining divergence could then be absorbed by the
counterterms $\delta c_2$ of the $\bmp^2/m^2$-currents to the left and to the
right of the current correlator.

The ambiguity in the renormalization procedure for the contribution
(i+ii) is related to the freedom of performing field redefinitions and does not
affect physical observables like the running of $c_1$. The contribution from (i+ii) to the $c_1$ evolution equation Eq.~\eqref{c1anomdim}
is independent of whether we use the 
counterterm $\delta {\cal V}_r$ or $\delta {\cal V}_{k3}$ to subtract the
relevant divergences. As we will see below, $B_\bmp$ and $B_3$ enter the final
expression for the NNLL anomalous dimension of $c_1$ in the combination $C_F
B_\bmp + B_3$ (see footnote~\ref{altfootnote}). Thus both approaches lead to the same result.
Together with Eq.~\eqref{Bi}, Eq.~\eqref{Bpshift} and \eqref{B3} represent the
main results of this work.

\section{Summary of Results}
\label{secresults}

From Eqs.~\eqref{RenDef}, \eqref{GendeltaVki} and~\eqref{epscheck} we obtain the anomalous dimensions
\begin{eqnarray}
\dn {\cal V}_{ki} &=& -\,\dn \delta {\cal V}_{ki} \,=\,
 2\,A_i\,(2\au-\as) + 4\,B_i\,(2 \au^2-\as^2) 
\,.
\label{Gen2loopRGE}
\end{eqnarray}
To integrate Eq.~\eqref{Gen2loopRGE} we can assume NLL running of $\au$ and
$\as$ in the $A_i$ and LL running in the $B_i$ terms. 
Respecting the matching conditions ${\cal V}_{ki}(\nu\!=\!1) = 0$, the solution
of Eq.~\eqref{Gen2loopRGE} at NLL order precision reads
\begin{eqnarray}
{\cal V}_{ki}^{\rm NLL}(\nu) &=& -\frac{4 \pi}{\beta_0} \,A_i\, \ln \frac{\au}{\as} \,+\, \left( \frac{\beta_1}{\beta_0^2}\,A_i - \frac{8 \pi}{\beta_0}\,B_i \right) \big( \au - \as \big)\,,
\label{Vki}
\end{eqnarray}
where 
$\beta_0 = \frac{11}{3}C_A - \frac43 T n_f \,,\; 
\beta_1 = \frac{34}{3}C_A^2 - 4 C_F T n_f - \frac{20}{3}C_A T n_f$ are the
standard one- and two-loop coefficients of the QCD beta
function. Eq.~\eqref{Vki} is simple, because it only accounts for purely
ultrasoft effects (plus the corresponding pull-up terms).
We recall that the coefficients ${\cal V}_{ki}^{(s)}$ contribute to the NLL anomalous dimension of $c_1$ with an additional factor $[{\cal V}_c^{(T),\rm LL}(\nu)]^2$ according to the definition of the ${\cal O}_{ki}$ sum operators.

The corresponding expressions for the $1/m^2$~potentials are given by~\cite{V2Vr}
\begin{align}
\big( {\cal V}_2^{(s)}(\nu) \big)_{us}^{\rm NLL} &=  {\cal V}_c^{(T),\rm LL}(\nu) \left[ - \frac{4 \pi}{\beta_0} A_{\bmk}^{(s)}  \ln \frac{\au}{\as} +  \left( \frac{ \beta_1}{\beta_0^2} A_{\bmk}^{(s)} - \frac{8 \pi}{\beta_0} B_{\bmk}^{(s)} \right) \big( \au \!-\! \as \big) \right] \,,
\label{V2NLL}
\\
\big( {\cal V}_r^{(s)}(\nu) \big)_{us}^{\rm NLL} &=  2 {\cal V}_c^{(T),\rm LL}(\nu) \left[ - \frac{4 \pi}{\beta_0} A_{\bmp}^{(s)}  \ln \frac{\au}{\as} +  \left( \frac{ \beta_1}{\beta_0^2} A_{\bmp}^{(s)} - \frac{8 \pi}{\beta_0} B_{\bmp}^{(s)} \right) \big( \au \!-\! \as \big) \right] \,.
\label{VrNLL}
\end{align}
In this work the soft Coulomb factor ${\cal V}_c^{(T),\rm LL}(\nu)$ has not been included in the $1/m^2$ potential operator structure in Eq.~\eqref{pots}, but appears directly in the Wilson coefficients ${\cal V}_{2,r}$ in Eqs.~\eqref{V2NLL} and~\eqref{VrNLL}. The square brackets have
the same form as Eq.~\eqref{Vki} and represent the ultrasoft
contributions (plus the corresponding pull-up terms) up to NLL order. In
addition to Eqs.~\eqref{V2NLL} and~\eqref{VrNLL} there are also purely soft
contributions to ${\cal V}_{2,r}$. They are fully known at LL order and can be
found in Ref.~\cite{amis}. 

Adopting the convention in Eq.~\eqref{Bpshift}, i.e. adding the extra contribution $\Delta B_\bmp^{(s)}$ to the singlet coefficient $B_\bmp^{(s)}$ determined in Ref.~\cite{V2Vr}\footnote{For the results in Ref.~\cite{V2Vr} the scheme choice of Eq.~\eqref{B3} was employed.} instead of introducing an additional coefficient $B_3^{(s)}$, we can summarize the coefficients in Eqs.~\eqref{Vki}-\eqref{VrNLL} as follows:\\
\begin{align}
 \bigg[\begin{array}{c} \! A_{\bmk}^{(s)} \! \cr \! B_{\bmk}^{(s)} \! \end{array}\bigg]
 &= C_F (C_A-2 C_F)  \bigg[\begin{array}{c} \! A^{(s)} \! \cr \! B^{(s)} \! \end{array}\bigg]\,, \nn\\
 \bigg[\begin{array}{c} \! A_{\bmp}^{(s)} \! \cr \! B_{\bmp}^{(s)} \! \end{array}\bigg] 
 &= -C_A C_F   \bigg[\begin{array}{c} \! A^{(s)} \! \cr \! B^{(s)} \! \end{array}\bigg] \,, \nn\\
\bigg[\begin{array}{c} \! A_1^{(s)} \! \cr \! B_1^{(s)} \! \end{array}\bigg] 
&= -C_A C_F (C_A-2 C_F) \bigg[\begin{array}{c} \! A^{(s)} \! \cr \! B^{(s)} \! \end{array}\bigg] \,, \nn\\
\bigg[\begin{array}{c} \! A_2^{(s)} \! \cr \! B_2^{(s)} \! \end{array}\bigg] 
 &= C_A C_F (C_A-4 C_F) \bigg[\begin{array}{c} \! A^{(s)} \! \cr \! B^{(s)} \! \end{array}\bigg] \,,
\label{AsAndBs}
\end{align}
where
\begin{align}
A^{(s)}= \frac1{3\pi}   \quad 
\text{\rm and} \quad B^{(s)}= \frac{C_A(47 + 6\pi^2) -10 n_f T }{108 \pi ^2} \,.
\end{align}
This simple form is quite suggestive because the ultrasoft corrections to the color singlet Wilson coefficients of the $1/m|\bmk|$ and the $1/m^2$ potentials are universal up to color factors. This appears to be a quite non-trivial fact given the complexity and differences in the computations for the $1/m^2$ and $1/m|\bmk|$ potentials. We emphasize again, however, that the form of Eq.~\eqref{AsAndBs} is depending on the scheme choice associated with Eq.~\eqref{Bpshift}.

Using Eqs.~\eqref{4qVk} and~\eqref{intfs} we can now write down the final
expression for the ultrasoft contributions 
(including soft mixing and pull-up terms) to the effective color singlet
$V_k$ potential in Eq.~\eqref{Vkeffinc1} up to NLL order.
The Wilson coefficient
\begin{align}
\lefteqn{\big({\cal V}_{k,{\rm eff}}^{(s)}(\nu)\big)_{us}^{\rm NLL}  = 
\frac{ [{\cal V}_c^{(T),\rm LL}(\nu) ]^2}{8 \pi^2}\, \big[\, 3 {\cal V}_{k1}^{(s)}(\nu) +2 {\cal V}_{k2}^{(1)}(\nu) 
\, \big]^{\rm NLL}}\qquad& \label{VkeffNLLsing}\\
&= -C_A C_F (C_A+2C_F) \frac{ [{\cal V}_c^{(T),\rm LL}(\nu) ]^2}{8 \pi^2}
\left[ - \frac{4 \pi}{\beta_0} A^{(s)}  \ln \frac{\au}{\as} +  \left( \frac{ \beta_1}{\beta_0^2} A^{(s)} - \frac{8 \pi}{\beta_0} B^{(s)} \right) \big( \au \!-\! \as \big) \right] \nn
\end{align}
appears in the NLL anomalous
dimension of the Wilson coefficient $c_1(\nu)$ of the ${}^3S_1$ heavy quark
current in Eq.~(\ref{c1anomdim}).
The LL running of the Coulomb coefficient ${\cal V}^{T}_c(\nu)$ is purely soft and
given by ${\cal V}^{(T),\rm LL}_c(\nu)= 4 \pi \as$. 
Together with the ultrasoft NLL running of the $1/m^2$ potential
coefficients ${\cal V}_{2,r}$ in Eqs.~\eqref{V2NLL} and~\eqref{VrNLL}, Eq.~\eqref{VkeffNLLsing} constitutes the NNLL ultrasoft mixing
contributions to the anomalous dimension of $c_1$.
\\

It is now straightforward to derive from Eq.~(\ref{c1anomdim}) 
the complete two-loop ultrasoft part of the NNLL mixing contributions to the
running of the ${}^3S_1$ current coefficient $c_1$. We can parametrize the form
of $c_1$ as~\cite{3loop}
\begin{eqnarray}
 \ln\Big[ \frac{c_1(\nu)}{c_1(1)} \Big] & = &
\xi^{\rm NLL}(\nu) + 
\Big(\,
\xi^{\rm NNLL}_{\rm m}(\nu) + \xi^{\rm NNLL}_{\rm nm}(\nu)
\,\Big) + \ldots
\,,
\label{c1solution}
\end{eqnarray}
where $\xi^{\rm NLL}$ is the well known NLL order
contribution~\cite{Pineda:2001et,HoangStewartultra} and
$\xi^{\rm NNLL}_{nm}$ is the NNLL non-mixing contribution determined in
Ref.~\cite{3loop}. The ultrasoft part of the NNLL mixing contribution $\xi^{\rm NNLL}_m$ is new and reads
\begin{align}
\xi^{\rm NNLL}_{\rm m, usoft} & = 
\frac{2\pi \beta_1}{\beta_0^3}\,\tilde A\,\alpha_s^2(m)\,
 \bigg[ -\frac{7}{4}+\frac{\pi^2}{6}+z\left(1-\ln\frac{z}{2-z}\right)
      +z^2\left(\frac{3}{4}-\frac{1}{2}\ln z\right)
\nn\\[1.5 ex] & \hspace{16 ex} 
      -\ln^2\left(\frac{z}{2}\right)+\ln^2\left(\frac{z}{2-z}\right)
      -2\mbox{Li}_2\left(\frac{z}{2}\right)\bigg]
\nn\\[1.5 ex] & 
+\,\frac{8\pi^2}{\beta_0^2}\,\tilde B\,\alpha_s^2(m)\,
  \Big[ 3-2z-z^2-4\ln(2-z) \Big]
\,,
\label{XiNNLLmix}
\end{align}
where\footnote{If one uses the counterterm $\delta {\cal V}_{k3}$ to absorb the
divergent contribution i+ii, the RHS of  Eq.~\eqref{VkeffNLLsing} reads
$
[{\cal V}_c^{(T),\rm LL}(\nu) ]^2/(8\pi^2)
[3 {\cal V}_{k1}^{(s)}(\nu) +2 {\cal V}_{k2}^{(1)}(\nu) 
+ 2{\cal V}_{k3}^{(s)}(\nu)]^{\rm NLL}$
and
$\tilde B =  C_F (B_\bmk^{(s)} +  2 B_\bmp^{(s)})
+ 3 B_1^{(s)} + 2 B_2^{(s)} + 2 B_3^{(s)} = -C_F(C_A+C_F)(C_A+2C_F) B^{(s)}$.
\label{altfootnote}
}
\begin{align}
\tilde A &=  C_F (A_\bmk^{(s)} +  2 A_\bmp^{(s)} )+ 3 A_1^{(s)} + 2 A_2^{(s)} = -C_F(C_A+C_F)(C_A+2C_F) A^{(s)}
 \,, \label{ASchlange} \\[1.5 ex]
\tilde B &=  C_F (B_\bmk^{(s)} +  2 B_\bmp^{(s)})+ 3 B_1^{(s)} + 2 B_2^{(s)} 
 = -C_F(C_A+C_F)(C_A+2C_F) B^{(s)} \,, \label{BSchlange} 
\end{align}
with
\begin{align}
z & \equiv  \left(\frac{\alpha_s(m\nu)}{\alpha_s(m)}\right)^{\rm LL}
  \, = \,
  \bigg(1+\frac{\alpha_s(m)\beta_0}{2\pi}\ln\nu\bigg)^{-1}
\,. \label{zLL}
\end{align}

\section{Numerical Discussion}
\label{sectiondiscussion}

\begin{figure}[htp]
\begin{center}
\includegraphics[width=\textwidth]{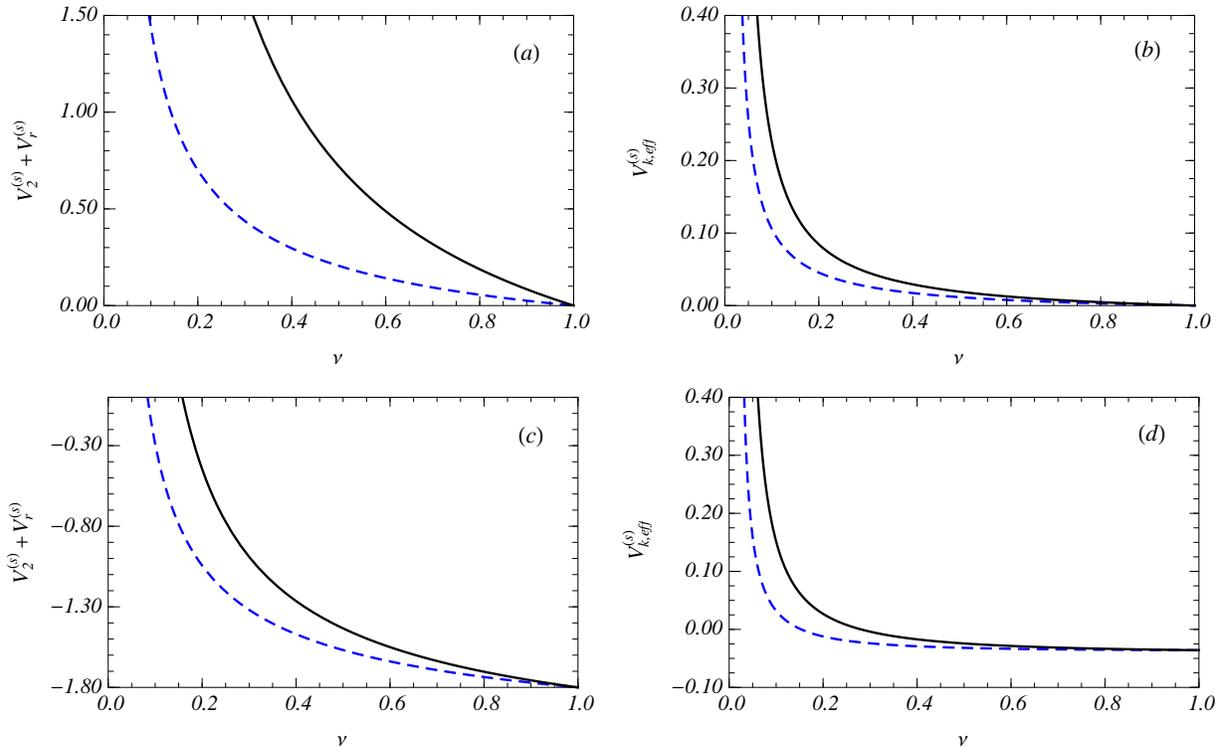}
\end{center}
\caption{Renormalization evolution in $\nu$ of  ${\cal V}_2^{(s)} + {\cal V}_r^{(s)}$ and ${\cal V}_{k,\rm eff}^{(s)}$. In panels~a and b the ultrasoft LL (dashed lines) and LL+NLL evolution (solid lines) are shown assuming zero matching conditions at $\nu=1$. 
The curves include, however, the soft LL running of the factors ${\cal V}_c^{(T)}$ and $[{\cal V}_c^{(T)}]^2$ and pull-up terms as in Eqs.~\eqref{V2NLL}, \eqref{VrNLL} and \eqref{VkeffNLLsing}. Panels c and d show the corresponding plots with the pure soft LL contributions and the corresponding matching conditions added to all curves.
 \label{VPlots}}
\end{figure}
In order to demonstrate the rather large size of the NLL order ultrasoft
corrections to the anomalous dimensions we plot in Fig.~\ref{VPlots} the evolution
of the coefficients of the $1/m^2$ potentials (${\cal V}_2^{(s)} + {\cal
  V}_r^{(s)}$) and of the $1/m|\bmk|$ potential ${\cal V}_{k,\rm eff}^{(s)}$  at LL
(dashed lines) and at NLL order (solid lines). These two sets of potentials are
directly affected by ultrasoft effects. Since they appear explicitly in the NLL
anomalous dimension of the ${}^3S_1$ current, they constitute
the most important ultrasoft mixing effects in the anomalous dimension of $c_1$,
see Eq.~\eqref{c1anomdim}.  
In panels a and b only the LL and NLL ultrasoft running (including soft
mixing and pull-up terms) according to Eqs.~\eqref{V2NLL}, \eqref{VrNLL} and \eqref{VkeffNLLsing} is displayed. For all curves zero matching conditions
at $\nu=1$ has been adopted. For both ${\cal V}_2^{(s)} + {\cal V}_r^{(s)}$ and
${\cal V}_{k,\rm eff}^{(s)}$ the ultrasoft NLL contributions are rather large
and can even exceed the LL ones. The situation remains in Figs.~\ref{VPlots}c and
d where we also include the full LL soft contributions and matching conditions from Ref.~\cite{amis}. 
Due to the small prefactor $\sim {\cal V}_c^{(s)}/(16 \pi^2)$, the contribution
from ${\cal V}_2^{(s)} + {\cal V}_r^{(s)}$ is relatively suppressed compared to
the effect of ${\cal V}_{k,\rm eff}$ in the evolution of $c_1$,
Eq.~\eqref{c1anomdim}. Therefore the ultrasoft NLL corrections to  ${\cal V}_{k,\rm eff}$ obtained in this
work have a substantially larger impact on $c_1$
than those for ${\cal V}_{2,r}^{(s)}$ obtained in Ref.~\cite{V2Vr}.

The interplay of the ultrasoft NLL evolution of the
potential coefficients ${\cal V}_2$, ${\cal V}_r$ and ${\cal V}_{k,\rm eff}$
in $c_1$ (see Eq.~(\ref{c1anomdim})) and the NNLL non-mixing contributions in the
running of $c_1$ is examined for the case of top and bottom quark pairs 
in table~\ref{tabcompare}, which represents an update to table~III of
Ref.~\cite{3loop}.  
\begin{table}[tbh]
\begin{center}
\begin{tabular}{|c||c|c|c||c|c|c|}
\hline
 & \multicolumn{3}{|c||}{$m=175$\,GeV}
 & \multicolumn{3}{|c|}{$m=4.8$\,GeV} \\ \hline 
 $\nu$ 
 & $\xi^{\rm NLL}(\nu)$ 
 & $\xi^{\rm NNLL}_{\rm nm}(\nu)$
 & $\xi^{\rm NNLL}_{\rm nm}(\nu)\!+\!\xi^{\rm NNLL}_{\rm m, us}(\nu)$
 & $\xi^{\rm NLL}(\nu)$ 
 & $\xi^{\rm NNLL}_{\rm nm}(\nu)$
 & $\xi^{\rm NNLL}_{\rm nm}(\nu)\!+\!\xi^{\rm NNLL}_{\rm m, us}(\nu)$
 \\ \hline\hline
1.0 & 0.0000 & 0.0000 & 0.0000 & 0.0000 & 0.0000 & 0.0000 \\ \hline
0.9 & 0.0031 & 0.0018 & 0.0018 & 0.0127 & 0.0158 & 0.0153 \\ \hline
0.8 & 0.0066 & 0.0040 & 0.0039 & 0.0267 & 0.0360 & 0.0335 \\ \hline
0.7 & 0.0105 & 0.0066 & 0.0063 & 0.0419 & 0.0629 & 0.0555 \\ \hline
0.6 & 0.0150 & 0.0099 & 0.0091 & 0.0582 & 0.1008 & 0.0824 \\ \hline
0.5 & 0.0200 & 0.0142 & 0.0126 & 0.0743 & 0.1585 & 0.1154 \\ \hline
0.4 & 0.0259 & 0.0202 & 0.0171 & 0.0852 & 0.2584 & 0.1530 \\ \hline
0.3 & 0.0328 & 0.0294 & 0.0230 & 0.0689 & 0.4789 & 0.1671 \\ \hline
0.2 & 0.0401 & 0.0460 & 0.0313 &  &  &  \\ \hline
0.1 & 0.0405 & 0.0916 & 0.0411 &  &  &  \\ \hline
\end{tabular}
\end{center}
{\caption{
Numerical values for the contributions to the running of $\ln c_1(\nu)/\ln c_1(1)$ as defined in Eq.~\eqref{c1solution}: $\xi^{\rm NLL}(\nu)$,
$\xi^{\rm NNLL}_{\rm nm}(\nu)$ and $\xi^{\rm NNLL}_{\rm m,us}(\nu)$. The values for $m$ are pole masses. 
The numbers are obtained by evaluation of the analytic results using 
one-loop running for $\alpha_s^{\rm LL}$ (in Eq.~\eqref{zLL}) and taking 
$\alpha_s^{(n_f=5)}(175\,\mbox{GeV})=0.107$ and 
$\alpha_s^{(n_f=4)}(4.8\,\mbox{GeV})=0.216$ 
as input.
}
\label{tabcompare} }
\end{table}
The values of $\ln c_1(\nu)/\ln c_1(1)$ at the NLL and NNLL level are 
given as a function of $\nu$ for the top quark ($m=175\,{\rm GeV}$) and the 
bottom quark ($m=4.8\,{\rm GeV}$) cases. The numbers in the third (sixth)
column show the sum of the complete (soft and ultrasoft) NNLL non-mixing and ultrasoft
mixing contributions. The latter terms were computed in this work. As a
comparison we display in the second (fifth) column the non-mixing contributions
alone. The first (fourth) column gives the number at the NLL order. As
anticipated, we see that there is a substantial cancellation between the
NNLL mixing and non-mixing contributions, particularly for small values of
$\nu\sim\alpha_s$, where the NRQCD matrix elements are being evaluated. 
This behavior is insofar remarkable as the NLL non-mixing contributions of $c_1$
arise from 3-loop vertex diagrams while the mixing corrections are
diagrammatically related to 4-loop vertex diagrams. This demonstrates in a quite
coercive way the importance of properly summing logarithms and of accounting for
mixing effects in NRQCD predictions. The relation of mixing and non-mixing
contributions to 3-loop and 4-loop vertex diagrams, respectively, also explains
why the mixing contributions are so much smaller than the non-mixing ones close
to the matching scale $\nu\sim 1$ where velocity logarithms are small. 

\begin{figure}[ht]
\begin{center}
\includegraphics[width=\textwidth]{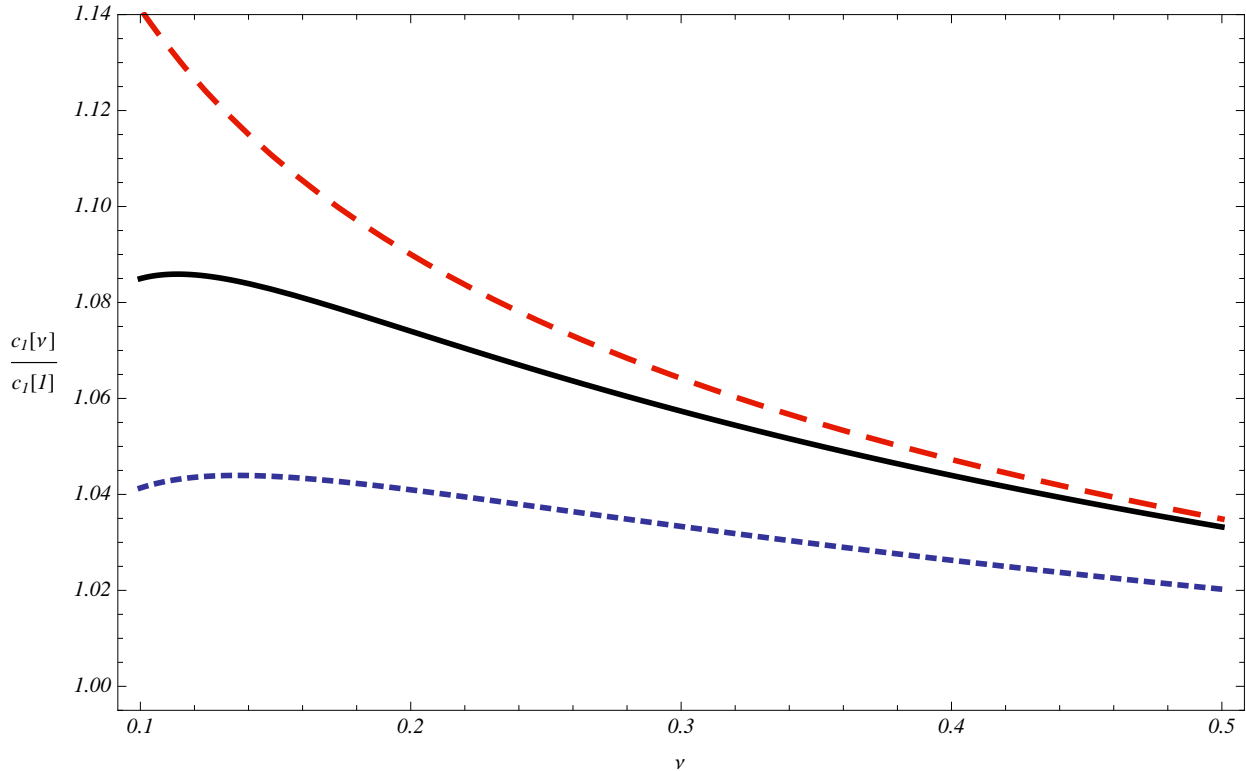}
\end{center}
\caption{Renormalization group evolution of the  ${}^3S_1$  current Wilson coefficient
$c_1(\nu)$ normalized to $c_1(1)$ for the top quark case ($m=175$~GeV). The
dotted (blue) line represents the full NLL result 
$\exp[\xi^{\rm NLL}]$, the dashed (red) line includes in addition the NNLL
non-mixing contributions, $\exp[\xi^{\rm NLL} + \xi_{\rm nm}^{\rm NNLL}]$. The
solid (black) line accounts for the full NLL and NNLL non-mixing contribution as
well as for the ultrasoft NNLL mixing corrections, $\exp[\xi^{\rm NLL} +
\xi_{\rm nm}^{\rm NNLL} +\xi^{\rm NNLL}_{\rm m, usoft}]$. The typical value for the
velocity scaling parameter relevant for the evaluation of top threshold matrix
elements is $\nu \sim \alpha_s\sim 0.15$. For the plot we have used $\alpha_s^{(n_f=5)}(175\,\mbox{GeV})=0.107$. \label{c1Plot}} 
\end{figure}

A visual display of the results is given in Fig.~\ref{c1Plot}, where the
evolution of $c_1(\nu)/c_1(1)$ is shown for top quark pair production close to
threshold ($m=175$~GeV). The blue dotted line represents the full NLL curve
known before from Refs.~\cite{Pineda:2001et,HoangStewartultra}. The red dashed
line includes in addition the NNLL non-mixing contributions computed in
Ref.~\cite{3loop} and exhibits a clear instability for $\nu\sim\alpha_s\sim
0.15$, where the low-energy matrix elements relevant for top threshold
production have to be evaluated. This instability is the cause of the relatively
 large 6-10\% uncertainty in the normalization of the
top pair threshold cross section~\cite{HoangEpiphany,PinedaSigner}. Finally, the
black solid curve shows $c_1(\nu)/c_1(1)$ including the full NLL and NNLL
non-mixing contributions as well as the ultrasoft NNLL mixing corrections
finalized in this work. The stabilization of the renormalization group evolution
for $\nu\sim\alpha_s\sim 0.15$ is obvious and will contribute to a reduction of
the current uncertainties in RGI predictions for
heavy quark pair threshold production. A detailed analysis of top and bottom
quark production will be the subject of future publications.

Nevertheless, one should not leave unmentioned that the NNLL corrections are
obviously still quite large. In particular, scale variations obtained from the
NLL order result around $\nu=0.15$ fail to give a reliable estimate of the
perturbative uncertainty. For heavy quark threshold production such behavior is
not an unfamiliar situation. For example, it is well known that the insertion of
${\cal O}(v^2)$ suppressed potentials and operators in NNLL order matrix
elements leads to large corrections that cannot be anticipated from scale
variation of the NLL results. In the case of the renormalization scale running
of $c_1(\nu)$ the rather large NNLL shift arises from the lower order logarithms
generated by the non-mixing contributions.

\section{Conclusion}
\label{sectionconclusion}

We have determined the two-loop ultrasoft corrections to the NLL order
renormalization group equation of the ${\cal O}(v)$ vNRQCD $1/m|\bmk|$
potentials.  
Combined with the previously known results for the ${\cal O}(v^2)$ potentials ${\cal V}_r^{(s)}(\nu)$ and ${\cal V}_2^{(s)}(\nu)$ determined in Ref.~\cite{V2Vr} our result completes the ultrasoft NNLL mixing part of the renormalization group evolution of the leading ${}^3S_1$ current that describes heavy quark-antiquark pair production at threshold. 
Together with the previously known NNLL non-mixing corrections our results present the parametrically dominant contributions of the NNLL anomalous dimension of the leading ${}^3S_1$ current and should allow for meaningful RGI predictions for heavy quark threshold production at the NNLL level.
\\ 

\noindent
Note added: 
While this paper was finalized we obtained a letter by Pineda~\cite{Pineda:2011aw}
on the NLL ultrasoft renormalization group running of the spin-independent
$\ord(1/m)$ and $\ord(1/m^2)$ potentials within the pNRQCD effective field
theory framework. Including the soft pull-up terms, see e.g. Eq.~\eqref{GendeltaVki}, we find agreement with our results, i.e. the previously known ultrasoft NLL results for the $1/m^2$ vNRQCD potentials $V_2^{(s)}$ and $V_r^{(s)}$~\cite{V2Vr} together with the $1/m|\bmk|$ potentials obtained in this work.
In particular our results in Eqs.~\eqref{V2NLL}, \eqref{VrNLL} and~\eqref{VkeffNLLsing} agree with the corresponding position space expressions $\delta V^{(2)}_{r,RG}(r;\nu_s,\nu_p)$, $\delta V^{(2)}_{p^2,RG}(r;\nu_s,\nu_p)$ and
$\delta V^{(1)}_{s,RG}(r;\nu_s,\nu_p)$ of Ref.~\cite{Pineda:2011aw}, respectively, upon transformation to momentum space and after imposing the correlation $\nu_p =
\nu_s^2/m \equiv \mu_U (= m \nu^2)$ between the pNRQCD soft matching scale $\nu_s$ and the pNRQCD
ultrasoft renormalization scale $\nu_p$. We remind the reader that the vNRQCD subtraction velocity $\nu$ is dimensionless whereas the pNRQCD scales $\nu_p$ and $\nu_s$ as well as the ultrasoft vNRQCD scale $\mu_U$ have the dimension of a mass.

The two-loop ultrasoft computations in our present work and in Refs.~\cite{V2Vr,PhD} were more involved than the corresponding ones in Ref.~\cite{Pineda:2011aw} using pNRQCD. This is related to the simpler power counting of pNRQCD where soft fluctuations are decoupled from potential and ultrasoft ones. In the vNRQCD formulation used in our work soft, potential and ultrasoft fluctuations are contained in the theory which allows to systematically treat higher order mixing effects involving all three types of fluctuations (see Ref.~\cite{3loop} for such a computation). The equivalence of the results obtained in Ref.~\cite{Pineda:2011aw} and the ones in this work (and in Refs.~\cite{V2Vr,PhD}) is technically non-trivial and indicates that the structural differences in the vNRQCD and pNRQCD formulations might not affect the description of purely ultrasoft effects.

\acknowledgments{
This work was supported in part by the EU network contract
MRTN-CT-2006-035482 (FLAVIAnet).
The Feynman diagrams in this paper have been drawn using {\tt JaxoDraw}~\cite{JaxoDraw}.
}

\appendix

\begin{section}{Feynman Rules of vNRQCD }
\label{feynmanrules}

In this Appendix we list the vNRQCD momentum space Feynman rules in Feynman gauge, which are relevant for our calculation.
Soft ($\pm\bmp$, $\pm\bmpp$) and ultrasoft momenta ($\vec k$, $\vec k_i$) of the heavy (anti)quarks are understood to point into the direction of the positive energy flow indicated by little arrows on the heavy quark lines. If not explicitly indicated in the pictures the default soft and ultrasoft momentum of a heavy (anti)quark is $\bmp$ and $\vec k$, respectively.
\\

\noindent
Propagators:
\begin{align}
\begin{array}{lcc}
\text{ultrasoft $A^0$ gluon:}&\hspace{5 ex} \raisebox{-1 ex}{\includegraphics[width = 0.15 \textwidth]{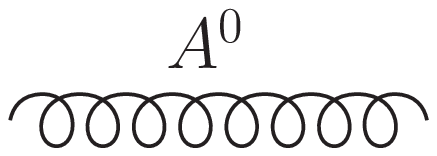}} \hspace{5 ex}& \displaystyle \frac{-i\,\delta^{AB}}{k^2+i\epsilon} \\[2 ex]
\text{ultrasoft $\bmA$ gluon:}&\hspace{5 ex} \raisebox{-1 ex}{\includegraphics[width = 0.15 \textwidth]{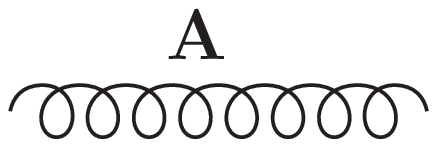}} \hspace{5 ex}& \displaystyle \frac{i\,\delta^{ij}\,\delta^{AB}}{k^2+i\epsilon} \\[2 ex]
\text{heavy (anti)quark:}&\hspace{5 ex} \raisebox{-1 ex}{\includegraphics[width = 0.15 \textwidth]{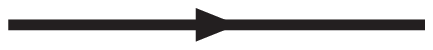}} \hspace{5 ex}& \displaystyle \frac{i}{k_0 - \frac{\bmp^2}{2m} + i\epsilon}
\end{array}
\end{align}

\noindent
Heavy quark kinetic insertions:
\begin{align}
\begin{array}{lc}
 \raisebox{-1 ex}{\includegraphics[width = 0.12 \textwidth]{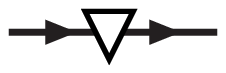}} \hspace{5 ex}& \displaystyle -i\, \frac{\bmp \cdot {\vec k}}{m} \\[2 ex]
 \raisebox{-1 ex}{\includegraphics[width = 0.12 \textwidth]{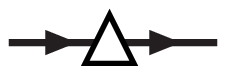}} \hspace{5 ex}& \displaystyle -i\, \frac{{\vec k}^2}{2m}
\end{array}
\end{align}

\noindent
Heavy quark - ultrasoft gluon - vertices\footnote{For the respective gluon-antiquark vertices replace $T^A \to {\bar T}^A$.}:
\begin{align}
\begin{array}{lc}
 \raisebox{-0.5 ex}{\includegraphics[width = 0.12 \textwidth]{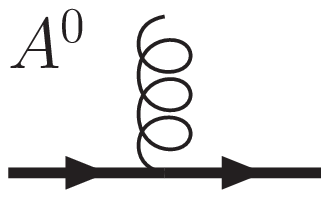}} \hspace{5 ex}& \displaystyle -i\, g\,T^A \\[1 ex]
 \raisebox{-0.5 ex}{\includegraphics[width = 0.12 \textwidth]{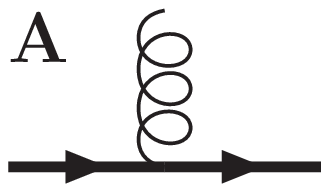}} \hspace{5 ex}& \displaystyle i\,g \frac{\bmp}{m}\,T^A
\end{array}
\end{align}

\noindent
Triple ultrasoft gluon - vertices\footnote{All gluon four-momenta $k_{A,B,C}^\mu$ are incoming.}:
\begin{align}
\begin{array}{lc}
 \raisebox{-5 ex}{\includegraphics[width = 0.17 \textwidth]{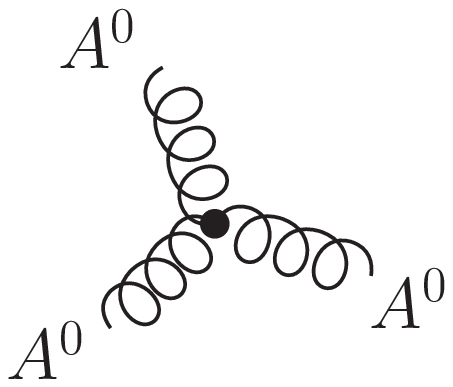}} \hspace{5 ex}& \displaystyle 0 \\[8 ex]
 \raisebox{-5 ex}{\includegraphics[width = 0.17 \textwidth]{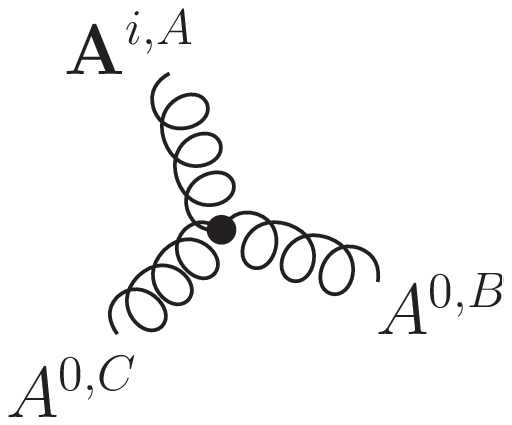}} \hspace{5 ex}& \displaystyle -g\,f^{ABC} (k_C^i-k_B^i) \\[8 ex]
 \raisebox{-5 ex}{\includegraphics[width = 0.17 \textwidth]{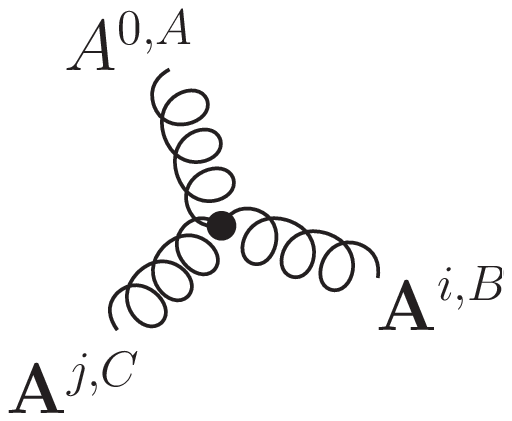}} \hspace{5 ex}& \displaystyle -g\,f^{ABC}\,\delta^{ij}\, (k_C^0-k_B^0)
\end{array}
\end{align}

\noindent
Heavy quark potential - vertices:
\begin{align}
\begin{array}{lc}
 \raisebox{-5 ex}{\includegraphics[width = 0.2 \textwidth]{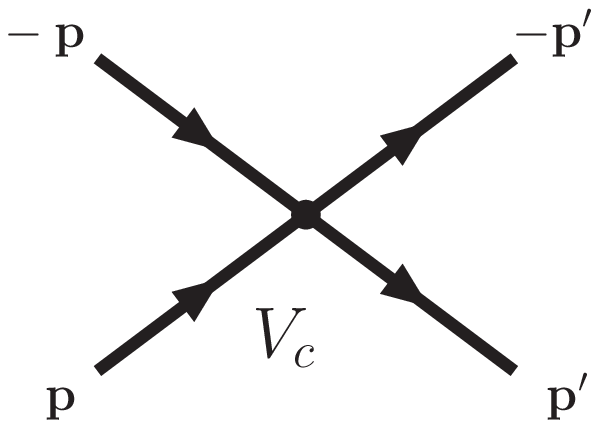}} \hspace{5 ex}& \displaystyle \frac{-i\, {\cal V}_c^{(T)}}{(\bmpp-\bmp)^2}\;\csT\\[5 ex]
%
\raisebox{-6 ex}{\includegraphics[width = 0.2 \textwidth]{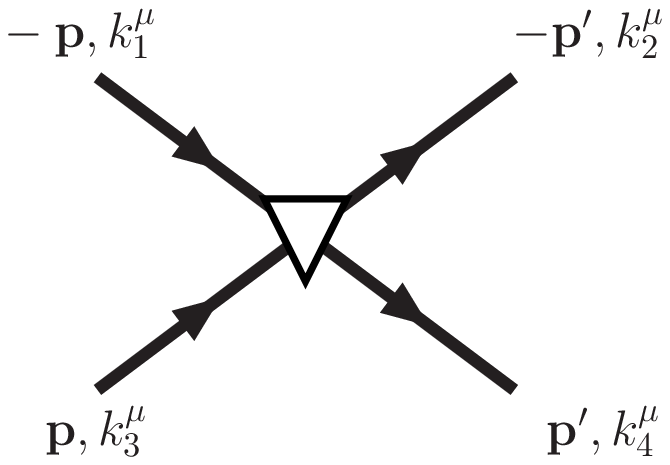}} \hspace{3 ex}& 
 i\, {\cal V}_c^{(T)}\, \csT\; \frac{(\bmpp-\bmp)\cdot [({\vec k_4}-{\vec k_3})-({\vec k_2}-{\vec k_1})]}{(\bmpp-\bmp)^4} \label{NablaPot}\\[5 ex]
\raisebox{-6 ex}{\includegraphics[width = 0.2 \textwidth]{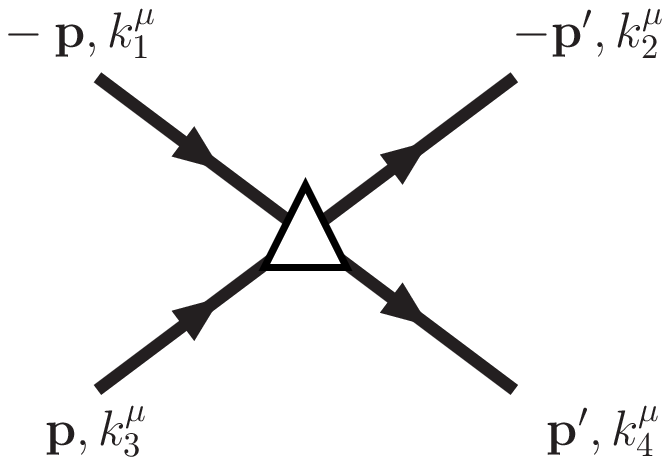}} \hspace{3 ex}& 
\begin{array}{l}
 i\, {\cal V}_c^{(T)}\, \csT\,\Big[ \frac{({\vec k_4}-{\vec k_3})^2+({\vec k_2}-{\vec k_1})^2}{2\,(\bmpp-\bmp)^4}   \\[2 ex]
\qquad- \frac{2[(\bmpp-\bmp)({\vec k_4}-{\vec k_3})]^2+2[(\bmpp-\bmp)({\vec k_2}-{\vec k_1})]^2}{(\bmpp-\bmp)^6} \Big]
\end{array}
\end{array}
\end{align}

\noindent
Heavy quark potential - ultrasoft gluon - vertex:
\begin{align}
\begin{array}{lc}
 \raisebox{-6 ex}{\includegraphics[width = 0.2 \textwidth]{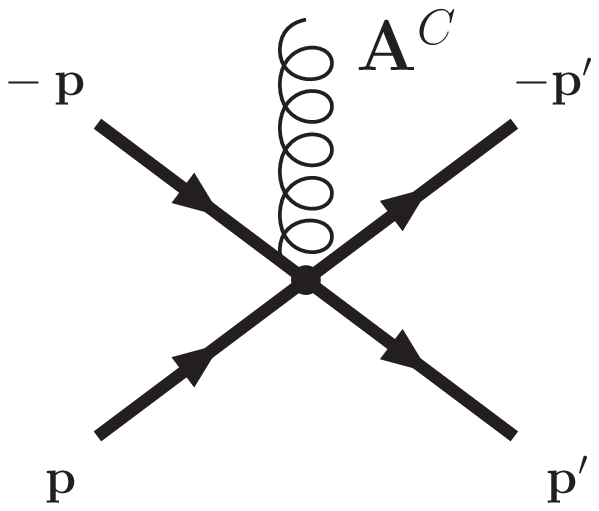}} \hspace{3 ex}& \displaystyle  -2\,g\, {\cal V}_c^{(T)}\,f^{ABC}\, T^A\!\otimes\!{\bar T}^B\, \frac{(\bmpp-\bmp)\cdot \bmA^C}{(\bmpp-\bmp)^4}
\end{array}
\label{AvecPot}
\end{align}

\end{section}


%

\end{document}